\documentclass[aps,superscriptaddress,twocolumn,pra,longbibliography,floatfix]{revtex4-2}
\usepackage{amsmath,amsthm,amsfonts,amssymb,bm,graphicx,color,mathpazo,times,braket}
\usepackage[colorlinks={true}, citecolor={blue}, filecolor={blue}, linkcolor={blue}, urlcolor={blue}]{hyperref}
\usepackage[caption=false]{subfig}
\usepackage{dcolumn} % Align table columns on decimal point
\usepackage{natbib} 
\newcommand{\abs}[1]{\left\vert#1\right\vert}

\begin{document}
%\preprint{APS/123-QED}
\title{Low-temperature quantum thermometry boosted by coherence generation}
\author{Asghar Ullah}
\email{aullah21@ku.edu.tr}
\affiliation{Department of Physics, Ko\c{c} University, 34450 Sar\i yer, Istanbul, T\"urkiye}
\author{M. Tahir Naseem} 
\email{mnaseem16@ku.edu.tr}
\affiliation{Department of Physics, Ko\c{c} University, 34450 Sar\i yer, Istanbul, T\"urkiye}
\affiliation{Faculty of Engineering Science, Ghulam Ishaq Khan Institute of Engineering Sciences and Technology, \\
Topi 23640, Khyber Pakhtunkhwa, Pakistan}
\author{\"Ozg\"ur E. M\"ustecapl\i o\u glu}	
\email{omustecap@ku.edu.tr}
\affiliation{Department of Physics, Ko\c{c} University, 34450 Sar\i yer, Istanbul, T\"urkiye}
\affiliation{T\"UBITAK Research Institute for Fundamental Sciences, 41470 Gebze, T\"urkiye}
\begin{abstract}
The precise measurement of low temperatures is significant for both the fundamental understanding of physical processes 
and technological applications. In this work, we present a method for low-temperature measurement that 
improves thermal range and sensitivity by generating quantum coherence in a thermometer probe. 
Typically, in temperature measurements, the probes thermalize with the sample being measured. However, 
we use a two-level quantum system, or qubit, as our probe and prevent direct probe access to the sample 
by introducing a set of ancilla qubits as an interface. We describe the open system dynamics of the probe 
using a global master equation and demonstrate that while the ancilla-probe system thermalizes with the sample, 
the probe \textit{per se} evolves into a nonthermal steady state due to nonlocal dissipation channels. 
The populations and coherences of this steady state depend on the sample temperature, allowing for precise and 
wide-range low-temperature estimation.  We characterize the thermometric performance of the method using quantum 
Fisher information and show that the quantum Fisher information can exhibit multiple and higher peaks at different 
low temperatures with increasing quantum coherence and the number of ancilla qubits.  Our analysis reveals that the 
proposed approach, using a nonthermal qubit thermometer probe with temperature-dependent quantum coherence generated 
by a multiple qubit interface between a thermal sample and the probe qubit, can enhance the sensitivity of 
temperature estimation and broaden the measurable low-temperature range.
\end{abstract}	

\maketitle

%--------------------------------------------------------------------------------%
\section{Introduction}\label{sec:intro}
%--------------------------------------------------------------------------------%

The achievement of precise measurements of a broad range of low temperatures is a key challenge in implementing quantum technologies~\cite{Mehboudi_2019,giovannetti2011advances,PhysRevLett.96.010401,toth2014quantum,celi2016quantum,RevModPhys.80.885,bloch2012quantum}. 
Quantum thermometry is a promising field of quantum metrology that can provide solutions for low-temperature measurements through the use of 
quantum thermometers~\cite{de2016local,PhysRevLett.122.030403,PhysRevB.98.045101,PhysRevX.10.041054,olf2015thermometry,PhysRevLett.96.130404,PhysRevA.103.023317,PhysRevLett.119.090603,PhysRevLett.125.080402,latune2020collective,global_QT,potts2019fundamental,
mehboudi2015thermometry,PhysRevA.82.011611, Zhang2022,Roman2019}. 
An ideal quantum thermometer should be significantly smaller than the sample being measured and its coupling to the sample should not alter its temperature.  Recent attention has been focused on 
quantum two-level systems, or qubits, as the smallest possible thermometer~\cite{PhysRevA.86.012125,PhysRevA.84.032105,bloch2012quantum,PhysRevA.91.012331,PhysRevLett.118.130502,PhysRevA.98.042124,PhysRevA.99.062114,PhysRevResearch.2.033497, PhysRevResearch.2.033394}, which has been demonstrated for ultracold gases~\cite{PhysRevX.10.011018}.  Increasing the dimensionality of the probe does not significantly 
improve the thermometer's thermal sensitivity~\cite{campbell2018precision}. In fact, an $N$-level optimal probe at thermal 
equilibrium must have $N-1$-fold degeneracy in the excited state, effectively reducing it to a two-level system~\cite{PhysRevLett.114.220405}. 
Such an optimal probe exhibits high precision estimation only for a specific temperature, indicated by a single peak in the quantum Fisher information.
 	
One of the primary challenges in thermometry is to find a specific probe that can accurately measure a broad range of low temperatures~\cite{Mehboudi_2019, global_QT,Mok2021, PhysRevA.104.052214,PhysRevA.105.042601,PhysRevA.105.012212,PhysRevLett.128.130502,PRXQuantum.3.040330}. 
Strong system-bath coupling can enhance a probe's range of thermal sensitivity, but the quantum Fisher information still exhibits a single 
peak, indicating that the probe remains optimal for only a single temperature~\cite{PhysRevA.96.062103, PhysRevB.98.045101,glatthard2022bending}. 
Achieving broader range thermal precision measurements requires extremely large degeneracy in the higher excited states 
of the probe~\cite{campbell2018precision}, which is difficult to achieve in physical implementations. Alternative proposals 
suggest applying external periodic control to the probe to obtain multiple peaks in the quantum Fisher information, but this 
comes at the cost of making the thermometer non-autonomous~\cite{Mukherjee2019,glatthard2022bending}. In a different approach, the global~\cite{global_QT,Mok2021} and Bayesian~\cite{PhysRevA.104.052214,PhysRevA.105.042601,PhysRevA.105.012212,PhysRevLett.128.130502,PRXQuantum.3.040330}  quantum thermometry schemes take into account the statistical properties of measurement outcomes and incorporate prior knowledge to broaden the thermal sensitivity range.
 	
Recent studies in quantum thermodynamics have provided a deeper understanding of the interplay between heat, work, quantum information, and specifically, quantum coherence~\cite{e15062100,binder2018thermodynamics,campbellbook,AsliReview,kurizki_kofman_2022}. In line with 
these quantum thermodynamical perspectives, several schemes have been proposed in quantum thermometry that exploit nonclassical features 
to enhance thermal sensitivity~\cite{PhysRevA.101.032112,PhysRevLett.125.080402,PhysRevLett.123.180602,PhysRevA.96.062103,potts2019fundamental,
	Campbell2017,PhysRevA.96.012316,PhysRevA.99.062114,PhysRevA.91.012331,PhysRevLett.118.130502,PhysRevA.98.042124,PhysRevResearch.2.033497,
	PhysRevA.98.050101}. However, most of these proposals require a dynamical approach, where temperature is measured 
in the transient regime~\cite{PhysRevA.98.042124,PhysRevResearch.2.033497}. This is because the probe eventually thermalizes in the steady-state, 
which is a diagonal (Gibbs) state devoid of coherence, resulting in the loss of all nonclassical features and quantum advantages. Moreover, 
for a probe in thermal equilibrium, temperature information is encoded solely in the populations, and quantum Fisher information equals 
classical Fisher information~\cite{PhysRevLett.114.220405, Paris_2015}. Previous studies have also not taken coherence or entanglement 
into account when considering the enhancement of the thermal sensitivity range, resulting in only a single peak being observed in 
the quantum Fisher information~\cite{PhysRevA.91.012331,PhysRevLett.118.130502,PhysRevA.98.042124,PhysRevResearch.2.033497}. 		
 		
We propose a method for precise and wide-range low-temperature measurement utilizing a multiqubit system thermalizing with a 
sample in thermal equilibrium. Our scheme measures temperature on a single probe qubit, taking advantage of its quantum coherence. 
The probe qubit is isolated from the sample and only interacts with the ancilla qubits, which bridge the probe and sample, generating 
coherence in the probe qubit in the steady state based on the sample's temperature. We analyze the open system dynamics of the 
ancilla-probe system connected to the thermal sample, deriving a global Markovian master equation that is consistent with the laws of 
thermodynamics in the Secular approximation~\cite{Levy_2014}. The nonlocal jump operators in this global master equation allow the sample 
to indirectly influence the probe qubit, encoding temperature information into both the diagonal (population) and off-diagonal (coherence) 
elements of the probe's density matrix. By judicious selection of the system parameters, the temperature-dependent coherence gives 
rise to multiple peaks in the quantum Fisher information, allowing for high-precision estimation across a range of temperatures. 
This approach, which enhances the probe's thermal sensitivity at very low temperatures, is consistent with dynamical 
ancilla-assisted quantum thermometry for a single-peak enhanced quantum Fisher information~\cite{PhysRevA.98.042124}. It is worth noting that at thermal equilibrium, the thermal sensitivity of the combined system comprising the ancilla and the probe is necessarily greater than the probe alone. However, in the case of a many-body probe, it might be experimentally challenging to perform the energy measurements required to achieve the Cramér-Rao bound~\cite{Mehboudi_2019}. In such scenarios, our proposed scheme, which exclusively relies on measurements performed over a single-qubit probe, can be useful, particularly in scaling problems that demand a many-body quantum system for probing the temperature~\cite{Zhang2022}. Additionally, a prominent feature of our scheme is that the reduced state of the probe in the local basis, obtained after tracing out the ancilla qubits, is a nonthermal state, which assists in enhancing the thermal sensitivity range.
 
The rest of the paper is organized as follows: In Sec.~\ref{sec:model}, we introduce our model of quantum thermometer and employ the global master equation approach to describe the open system dynamics of the model. In addition, we briefly discuss quantum Fisher information (QFI), a figure of merit in quantum metrology (Sec.~\ref{sec QFI}).
In Sec.~\ref{sec results}, we present and discuss the QFI for the case of single and two ancilla qubits. Furthermore, we discuss the case when the probe has direct access to the sample, such as ancilla qubits in Sec.~\ref{sec global}. We discuss the results for all the thermalized qubits and describe the two cases for identical and nonidentical qubits in Secs.~\ref{sec ident} and~\ref{non ident}, respectively.
The concluding remarks of this study are given in Sec.~\ref{sec conc}.
	
%--------------------------------------------------------------------------------%
\section{Model and Preliminaries}\label{sec:II}
%--------------------------------------------------------------------------------%
In this section, we introduce our theoretical model, which can boost the thermal sensitivity of 
a two-level system probe at low temperatures via coherence generation. 
Then, we briefly discuss QFI, a key figure of merit for the thermal sensitivity of a quantum thermometer. 

%--------------------------------------------------------------------------------%
\subsection{Model description}\label{sec:model}
%--------------------------------------------------------------------------------%
%-------------------------------------------------------------%
\begin{figure}[t]
  	\centering
  	\subfloat{
  		\includegraphics[scale=0.37]{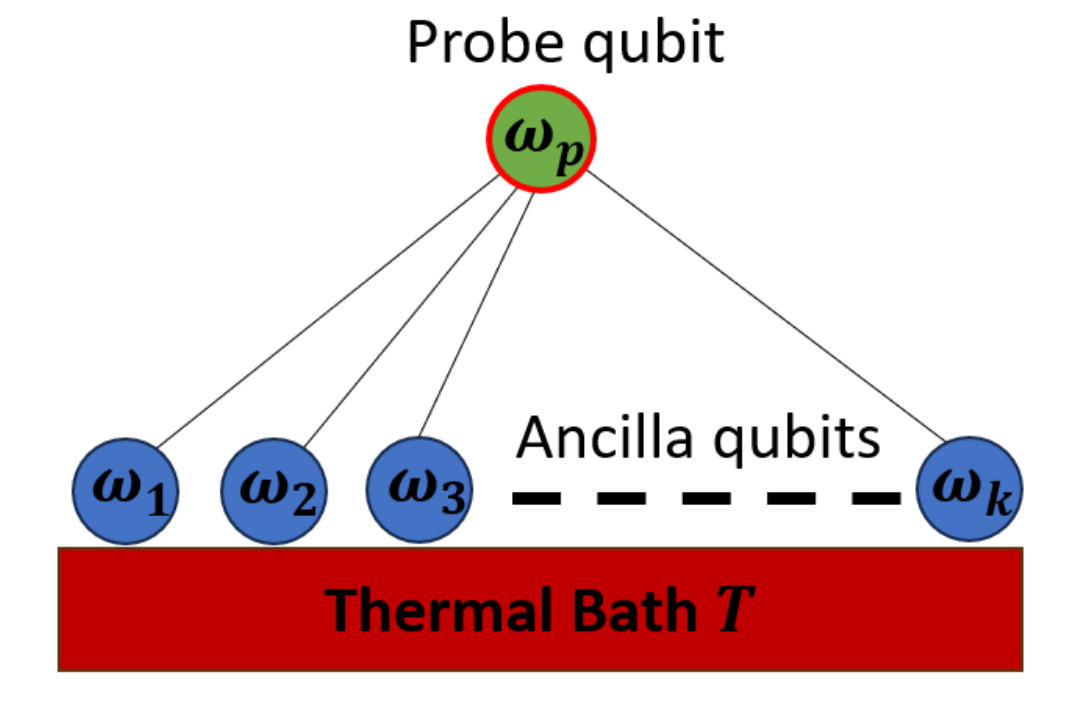}}
  		\caption{A schematic illustration of our quantum thermometry model. The ancilla qubits with transition frequencies $\omega_k$ are 
  		coupled to a thermal bath (sample) at temperature $T$. The probe qubit with transition frequency $\omega_p$ has no direct access to 
  		the thermal bath; instead, it is coupled with the ancilla qubits and indirectly probes the temperature of the thermal sample.
 		The probe qubit, lying outside the bath, is used for the estimation of $T$, and the energy measurements 
		are only performed over the probe qubit.}\label{fig1}
\end{figure}
%--------------------------------------------------------------%
We consider a qubit probe of transition frequency $\omega_{p}$ coupled to $N$ ancilla qubits of 
frequency $\omega_{k}$ ($k= 1, 2,..., N$). The ancilla qubits are coupled directly to a thermal 
bath (thermal sample) of temperature $T$ whose temperature is to be estimated. The qubit probe is 
outside the thermal sample and does not have direct access to it, as shown in Fig.~\ref{fig1}. 
The Hamiltonian of the ancilla-probe system is given by~\cite{PhysRevA.98.042124, PhysRevE.99.042121, 
	Naseem_2020, PhysRevResearch.2.033285} 
(we take $\hbar=1$)
\begin{equation}\label{eq1}
  \hat{H}=\frac{1}{2}\omega_p\hat{\sigma}^z_p+\sum\limits_{k=1}^{N}\frac{1}{2}\omega_k\hat{\sigma}^z_k
  +\sum\limits_{k=1}^{N}g_k\hat{\sigma}^z_k\hat{\sigma}^x_p,
\end{equation}
where $g_k$ is the coupling strengths between the ancilla and probe, and
$\hat{\sigma}^\zeta_{j}$ ($j=p, k$) with $\zeta_=x,y,z$ are the Pauli matrices. In Eq.~(\ref{eq1}), the 
first, second, and third terms denote the Hamiltonians of the probe, ancilla, and their interaction, respectively. 
We note that the asymmetric interaction between the qubits given in Eq.~(\ref{eq1}) has previously been employed 
for the proposals of ancilla-assisted thermometry~\cite{PhysRevA.98.042124}, quantum thermal diode~\cite{PhysRevE.99.042121}, 
quantum absorption refrigerator~\cite{Naseem_2020}, and quantum heat manager~\cite{PhysRevResearch.2.033285}. 
Recently, Kolář \textit{et al.} considered a similar asymmetric interaction between two-level
systems where they showed that the quantum coherence appears by a different mechanism, whereas the system-bath coupling 
does not have to be engineered~\cite{kolavr2022achieving}. This is done by formulating a necessary condition 
for the autonomous generation of non-zero local coherence in the individual two-level system (TLS).
  
We consider asymmetric interaction between the ancilla and probe qubits, which leads to temperature-dependent coherences in 
the probe [see Eq.~(\ref{eq:coherences})]. Such temperature-dependent coherences lead to the enhancement of the thermal 
sensitivity of the probe (Fig.~\ref{fig2}). We note that initial coherences or coherent driving on the probe to induce coherence 
do not enhance the thermal sensitivity of the probe because the coherences, in this case, are 
temperature independent~\cite{PhysRevLett.114.220405}. 
In addition, other symmetric or anti-symmetric interactions between the ancilla and probe qubits may not generate coherence in 
the probe. For instance, we have shown in Appendices~\ref{app:A} and~\ref{app:B} that dipole-dipole interaction and Dzyaloshinskii-Moriya 
interaction do not generate any coherence in the probe qubit, respectively. For the implementation of the Hamiltonian given in Eq.~(\ref{eq1}), we provide some examples for the realization and possible directions for such asymmetric interactions in Appendix ~\ref{exp}. We emphasize that quantum coherence is not the only resource for 
enhancing thermal sensitivity, and it’s role on the thermal range and sensitivity can depend on the model; for instance, in our model, 
it is the asymmetric type of interaction between the probe and ancilla qubits that generates coherence in the probe's state, which widens 
the range and boosts the sensitivity of low-temperature measurements.
  
To investigate the dynamics of the ancilla-probe system, we employ the standard Born-Markov master equation~\cite{Breuer}. 
The derivation of the master equation for coupled systems requires diagonalization of the total system Hamiltonian. 
The Hamiltonian of the ancilla-probe system can be diagonalized using the unitary transformation~\cite{PhysRevE.99.042121},
\begin{equation}\label{eq:transformation}
U :=\exp[-\frac{i}{2}\sum\limits_{k=1}^{N}\theta_k\hat{\sigma}^z_k\hat{\sigma}^y_p],
\end{equation}
where the angle $\theta_k$ is defined as
\begin{equation}
    \theta_k:=\arctan(\frac{2g_k}{\omega_p}).
\end{equation}
This gives the diagonalized Hamiltonian in the form
\begin{equation}
\tilde{H}=\sum\limits_{k=1}^{N}\frac{\omega_k}{2}\tilde{\sigma}^z_k+\frac{\Omega}{2}\tilde{\sigma}^z_p.\label{eq:DiagHamil}
\end{equation}
The transformed frequency $\Omega$ of the probe-qubit depends on the number of ancilla qubits, and for a single and two ancilla qubits, 
the explicit expressions are given in Eqs.~(\ref{eq:singletrans}), and~(\ref{eq:omega2}), respectively. 
After the transformation, the Pauli matrices of the ancilla and probe qubits take the form
\begin{eqnarray}\label{eq:Paulitrans}
\tilde{\sigma}^x_k &=& \cos\theta_k\hat{\sigma}^x_k+\sin\theta_k\hat{\sigma}^y_k\hat{\sigma}^y_p, \nonumber \\
\tilde{\sigma}^y_k &=& \cos\theta_k\hat{\sigma}^y_p-\sin\theta_k\hat{\sigma}^x_k\hat{\sigma}^y_p, \quad\quad \tilde{\sigma}^z_k = \hat{\sigma}^z_k \nonumber \\
\tilde{\sigma}^x_p &=& \cos\theta_k\hat{\sigma}^x_p-\sin\theta_k\hat{\sigma}^z_k\hat{\sigma}^z_p, \quad\quad \tilde{\sigma}^y_p=\hat{\sigma}^y_p, \nonumber\\
\tilde{\sigma}^z_p &=&\cos\theta_k\hat{\sigma}^z_p+\sin\theta_k\hat{\sigma}^z_k\hat{\sigma}^x_p.
\end{eqnarray}	

As an example, we assume a bosonic thermal bath, whose temperature is to be probed, coupled to ancilla qubits. 
The interaction Hamiltonian of the ancilla and bath has the form
\begin{equation}\label{eq:int}
	\hat{H}_\text{int} = \sum_{l,k} g_{l,k}\hat{\sigma}^{x}_{k}(\hat{a}_{l}+\hat{a}^{\dagger}_{l}),
\end{equation}
where $\hat{a}_{l}$ ($\hat{a}^{\dagger}_{l}$) is the annihilation (creation) bosonic operator of the  
$l$th bath mode, and $g_{k,l}$ indicates the coupling strength between the $l$th bath mode with the $k$th ancilla qubit. The global master equation 
for the coupled ancilla-probe system after the Born-Markov and Secular approximations can be written 
as~\cite{PhysRevE.99.042121, Naseem_2020, PhysRevResearch.2.033285} (For notational simplicity, 
we use the summation convention to drop the summation sign such that summation over repeated index $k$ is implied.)
\begin{equation}
\begin{aligned}
\dot{\tilde{\rho}} =&-i[\tilde{H},\tilde{\rho}]+\cos^2\theta_k \mathcal{G}(\omega_k)\Big(\mathcal{D}[\tilde{\sigma}^-_k]+e^{-\omega_k/T}\mathcal{D}[\tilde{\sigma}^+_k]\Big)\\
& +\sin^2\theta_k \mathcal{G}(\omega_{-k})\Big(\mathcal{D}[\tilde{\sigma}^-_k\tilde{\sigma}^+_p]+e^{-\omega_{-k}/T}\mathcal{D}[\tilde{\sigma}^+_k\tilde{\sigma}^-_p]\Big)\\
&
+\sin^2\theta_k \mathcal{G}(\omega_{+k})\Big(\mathcal{D}[\tilde{\sigma}^-_k\tilde{\sigma}^-_p]+e^{-\omega_{+k}/T}\mathcal{D}[\tilde{\sigma}^+_k\tilde{\sigma}^+_p]\Big)\label{eq:master}.
\end{aligned}
\end{equation}
Here, $\omega_{\pm k} = \omega_{k}\pm\Omega$, and mixing angle $\theta_{k}$ depend on the number of ancilla qubits; 
for a single and two ancilla qubits $\theta_{k}$ is given in Eqs.~(\ref{eq:singletrans}), and ~(\ref{eq:angle2}), 
respectively. In Eq.~(\ref{eq:master}), $\mathcal{D}[\Tilde{c}]$ is the Lindblad superoperator and is defined as
\begin{equation}
\mathcal{D}[\tilde{c}]=\tilde{c}\tilde{\rho}\tilde{c}^\dagger-\frac{1}{2}(\tilde{c}^\dagger \tilde{c}\tilde{\rho}+\tilde{\rho}\tilde{c}^\dagger \tilde{c}).
\end{equation}
The raising (lowering) spin operators $\tilde{\sigma}^+(\tilde{\sigma}^-)$ are defined as 
\begin{equation}
\tilde{\sigma}^\pm_{k,p}=\frac{1}{2}(\tilde{\sigma}^x_{k,p}\pm i\tilde{\sigma}^y_{k,p}).
\end{equation}
The bath spectral response function is denoted by $\mathcal{G}(\omega_k)$ in Eq.~(\ref{eq:master}). 
An appealing feature of our scheme is that the thermal sensitivity of the probe does not depend on the bath 
spectrum. The ancilla-probe system evolves to a steady state independent of the bath spectrum (see Sec.~\ref{sec results} for details). 
In the master equation~(\ref{eq:master}), we have neglected cross dissipators of the form $\mathcal{D}[\tilde{\sigma}_{k},\tilde{\sigma}_{k^{\prime}}]$; this 
is well justified if the distance between the ancilla qubits is much larger compared to the wavelength of the bath 
quanta (for example, photons)~\cite{Damanet_2016,PhysRevA.99.052105,PhysRevE.102.042111}. 

%--------------------------------------------------------------------------------------------------------
\subsection{Quantum Fisher information} \label{sec QFI}
%--------------------------------------------------------------------------------------------------------

In quantum thermometry, one requires a quantum system for the estimation of the unknown temperature of a sample. 
A small thermometer is desired compared to the sample so that it should not significantly affect the temperature of the sample. In this respect, 
the smallest possible thermometer can be a two-level quantum system (such as a qubit) which is small yet efficient 
as a probe for estimation purposes~\cite{bloch2012quantum, PhysRevA.91.012331,PhysRevLett.118.130502,PhysRevA.98.042124,
	PhysRevA.99.062114,PhysRevResearch.2.033497, PhysRevResearch.2.033394}. The qubit probe initially prepared in a state $\hat{\rho}_p$ 
allows interaction with a bath whose temperature is to be estimated. The bath information is imprinted onto the probe state to be mapped 
via some estimator $\hat{T}$~\cite{paris2009quantum}. Fundamental bounds limit the precision of parameter estimation. 
The optimal estimators 
in quantum estimation theory are those saturating the quantum Cram\'er-Rao 
bound inequality such that~\cite{cramer1999mathematical,helstrom1969quantum,PhysRevLett.72.3439}
\begin{equation}
\Delta T^2\geq\frac{1}{\mathcal{N}\mathcal{F}_Q(T)}.
\end{equation}
Here, $\Delta T^2$ is the temperature variance and $\mathcal{N}$ denotes the number of measurements. 
$\mathcal{F}_Q(T)$ is the quantum Fisher information which quantifies the information about temperature 
$T$ encoded onto the state of the probe, and it is defined as
\begin{equation}
\mathcal{F}_Q(T)=\text{Tr}[\rho(T) L_\text{T}^2].
\end{equation}
$L_\text{T}$ is the symmetric logarithmic derivative satisfying the equation $2\partial_\text{T}\rho(T)=\{L_\text{T},\rho(T)\}$. 
As depicted in Fig.~\ref{fig1}, our quantum thermometer is a single qubit probe interacting indirectly with the thermal 
bath via an interface of ancilla qubits. Accordingly, we will only focus on the QFI of a single qubit, 
which is given by~\cite{dittmann1999explicit,PhysRevA.87.022337}
\begin{equation}\label{eq:TLSfisher}
\mathcal{F}_Q(\hat{\rho})=\text{Tr}\Big[\Big(\frac{\partial \hat{\rho}}{\partial T}\Big)^2\Big]+\frac{1}{\text{Det}(\hat{\rho})}\text{Tr}\Big[\Big(\hat{\rho}\frac{\partial\hat{\rho}}{\partial T}\Big)^2\Big].
\end{equation}
The mathematical basis of our scheme can be seen in this expression, which can be used for both thermal and nonthermal qubit states. 
If there are any bath temperature-dependent coherences in the probe qubit state, they can be used to induce multiple 
peaks and enhancements in QFI.

%-------------------------------------------------------------------------
\section{results}\label{sec results}
%-------------------------------------------------------------------------

In this section, we present results for the low-temperature quantum thermometry for the model described in 
Fig.~\ref{fig1} by considering one (Sec.~\ref{sec N1}) and 
two (Sec.~\ref{secN2}) ancilla qubits.
We show that multiple peaks in the quantum Fisher information can be obtained by harvesting quantum coherence in the probe induced by 
ancilla qubits.

%-------------------------------------------------------------------------------------------------
\subsection{Enhancement of low-T sensitivity with a single ancilla qubit ($N=1$)}\label{sec N1}
%-------------------------------------------------------------------------------------------------

In the case of a single ancilla qubit, there is no summation and $k=1$ in the master equation~(\ref{eq:master}). 
The diagonalized Hamiltonian of the ancilla-qubit system $\tilde{H}$ is given in Eq.~(\ref{eq:DiagHamil}). 
The transformed frequency of the probe qubit and the mixing angle $\theta_{1}$ are given by
\begin{equation}\label{eq:singletrans}
   \Omega = \sqrt{\omega^{2}_{p}+4g^2_{1}},\quad  \text{and}\quad\theta_{1} = \text{arctan}\bigg(\frac{2g_1}{\omega_p}\bigg),
\end{equation}
respectively. The second term in the master equation~(\ref{eq:master}) describes the energy exchange between the ancilla and the thermal bath 
via the local dissipation channel. The last two terms, on the contrary, correspond to the non-local dissipation 
channels between the bath and ancilla-probe system. These two terms show that the probe qubit can exchange energy with 
the bath via nonlocal dissipation processes, though it is not directly coupled with the bath. We emphasize that the Lindblad dissipators $\mathcal{D}[\tilde{\sigma}^-_k\tilde{\sigma}^-_p]$ and $\mathcal{D}[\tilde{\sigma}^+_k\tilde{\sigma}^+_p]$ in the master equation (7) are different from the squeezed-like dissipators. The nonlocal squeezed-like dissipators, e.g.,  $\mathcal{D}[\tilde{A},\tilde{A}]:= \tilde{A}\tilde\rho\tilde{A} - \frac{1}{2}(\tilde{A}\tilde{A}\tilde{\rho}+\tilde{\rho}\tilde{A}\tilde{A})$, where $\tilde{A}\in [\tilde{\sigma}^-_{k}\tilde{\sigma}^-_{p}, \tilde{\sigma}^+_{k}\tilde{\sigma}^+_{p}]$, if present in the master equation might generate entanglement between the ancilla and probe qubits. However, the dissipative terms representing dissipation in a common thermal squeezed-like bath are not present in the master equation (7). This absence is expected since, in our scheme, one of the coupled qubits weakly interacts with a single thermal bath in the absence of any external control drive. In our approach, we perform a microscopic derivation of the Lindblad Markovian master equation using the diagonalized system Hamiltonian, which is expressed in the dressed basis [Eq.~(\ref{eq:DiagHamil})]. This derivation leads to the equilibrium steady-state of the system  (in the dressed basis), represented by~\cite{Breuer}
\begin{equation}
\tilde{\rho}_\text{ss}=\tilde{\rho}_1\otimes\tilde{\rho}_p.
\end{equation}
Here $\tilde{\rho}_{1}$ and $\tilde{\rho}_{p}$ are the reduced states of the ancilla and probe qubits, which are given by
\begin{equation}
	\Tilde{\rho}_1=\frac{e^{-\frac{1}{2}\beta\omega_{1}\tilde{\sigma}^{z}_{1}}}{\mathcal{Z}_1},\quad\text{and}\quad \Tilde{\rho}_p
	=\frac{e^{-\frac{1}{2}\beta\Omega\tilde{\sigma}^{z}_{p}}}{\mathcal{Z}_p},
\end{equation}
respectively. Here, $\mathcal{Z}_1=\text{Tr}[\exp(-\beta\omega_{1}\tilde{\sigma}^{z}_{1}/2)]$ and $\mathcal{Z}_p=\text{Tr}[\exp(-\beta\Omega\tilde{\sigma}^{z}_{p}/2)]$ are the partition functions of the ancilla and probe qubits, respectively; and $\beta=1/k_BT$ is the inverse temperature. Henceforth, we take the Boltzmann constant $k_B=1$. 
The steady-state density matrix is written in the
eigenstates of the dressed Hamiltonian $\tilde{H}$ [for example, see Eq.~(\ref{eq:DiagHamil})], and these are given by the individual 
eigenstates of the ancilla and probe qubits as,
\begin{eqnarray}
	\ket{1} = \cos\frac{\theta_1}{2}\ket{++} - \sin\frac{\theta_1}{2}\ket{+-}, \nonumber \\
	\ket{2} = \sin\frac{\theta_1}{2}\ket{++} + \cos\frac{\theta_1}{2}\ket{+-}, \nonumber \\
	\ket{3} = \cos\frac{\theta_1}{2}\ket{-+} + \sin\frac{\theta_1}{2}\ket{--}, \nonumber \\
	\ket{4} = \cos\frac{\theta_1}{2}\ket{--} - \sin\frac{\theta_1}{2}\ket{-+}\label{eq:eigstates},
\end{eqnarray}
with their corresponding eigenvalues $\epsilon_1 = (\omega_{1} + \Omega)/2$,  $\epsilon_2 = 
(\omega_{1} - \Omega)/2$, $\epsilon_3 = (-\omega_{1} + \Omega)/2$, and $\epsilon_4 = (-\omega_{1} - \Omega)/2$, 
respectively. However, in the basis of the individual qubits, the steady state of the ancilla and probe qubits does not necessarily equal a product state $\hat{\rho}_\text{ss}\neq\hat{\rho}_1\otimes\hat{\rho}_p$.
	
%------------------------ Figure 2 ------------------------
\begin{figure}[t]
\centering
\subfloat[]{
\includegraphics[scale=0.65]{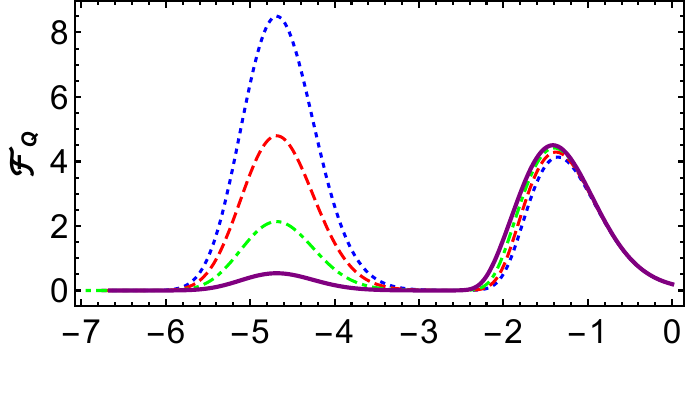}}\\
\subfloat[]{
\includegraphics[scale=0.63]{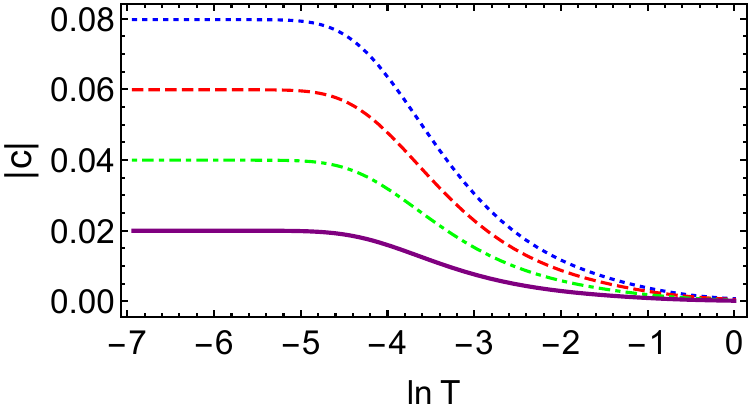}}
\caption{ \textbf{(a)} Quantum Fisher information $\mathcal{F}_{Q}$ associated with the reduced state of the probe qubit $\hat{\rho}_{p}$ by performing local measurements on the two-level system probe as a function of the temperature $T$ for different coupling strengths between the ancilla and probe qubits. \textbf{(b)} The absolute values of coherence $\abs{c}$ as a function of thermal bath temperature $T$ for different coupling strengths between the ancilla and probe qubits. The solid, dot-dashed, dashed, and dotted curves are for $g_{1}=0.01, 0.02, 0.03, 0.04$, respectively. In both panels, the other system parameters are the same: $\omega_p=1$ and $\omega_1 =0.04$. In the weak ancilla-probe coupling limit, the location of the two peaks in \textbf{(a)} can be determined by using Eq.~(\ref{eq:QFIapprox}), and given by $T^*_1=\gamma\omega_1$ and $T^*_p=\gamma\omega_p$. Here $\gamma$ is a constant and determined by Eq.~(\ref{eq:gamma}).}\label{fig2}
\end{figure}
 %-----------------------------------------------------------%

Our probe is a single qubit placed outside the thermal bath (Fig.~\ref{fig1}), 
and we wish to perform local measurements on this probe qubit. Therefore, 
by using the transformation given in Eq.~(\ref{eq:transformation}), the reduced state of 
the probe qubit in the local basis is given by
\begin{equation}
	\hat{\rho}_{p}= \text{Tr}_1[U\tilde{\rho}_\text{ss}U^\dagger]=\frac{1}{2}\begin{pmatrix}
	1-\chi & c\\
	 c & 1 +\chi
\end{pmatrix},\label{ProbD}
\end{equation}
where the diagonal (populations) and off-diagonal (coherences) terms are determined by
\begin{eqnarray}
\chi &=&\cos(\theta_1)\tanh(\frac{\Omega }{2 T}), \quad\text{and} \\
c &=& \sin(\theta_1)\tanh(\frac{\Omega}{2T})\tanh (\frac{\omega_1}{2 T}),
\label{eq:coherences}
\end{eqnarray}
respectively. In addition to the free Hamiltonian and coherent interaction parameters $\omega_p,\omega_1$ and $g_1$, through $\theta_1$ and $\Omega$, 
both the coherence and populations of the probe qubit 
carry information on the temperature of the sample. In contrast to thermalized probes with the sample,
such a non-thermal probe state allows us to manipulate QFI with additional terms associated with the temperature-dependent coherences. 
We shall explore different system parameter regimes to search for
multiple peaks and enhancement of thermal sensitivity effects in the low-temperature regimes in QFI,
computed by using Eq.~(\ref{eq:TLSfisher}). It is worth noting that the QFI is computed by considering the local 
measurements on our probe qubit. 

The complete expression for the QFI is too cumbersome to report here. 
We depict QFI $\mathcal{F}_{Q}$, and absolute coherence $\abs{c}$ in the probe as a function of the bath temperature 
$T$ for different values of coupling strength $g$ in Figs.~\ref{fig2}(a) and~\ref{fig2}(b), respectively. It is immediately evident that the 
lower peak in the QFI is associated with the coherence developed in the probe qubit. By increasing the ancilla-probe coupling strength, coherence 
in the probe qubit become larger, thus enhancing the probe's thermal sensitivity at lower temperatures. It is often useful to evaluate 
the relative error bound in estimating the temperature for the performance of a given probe~\cite{Mukherjee2019,glatthard2022bending}. 
We discuss the relative error bound in the case of our scheme in Appendix~\ref{app:relative}.

We emphasize that, for a two-level system probe at thermal equilibrium with the bath, it is not possible to 
have multiple peaks in the QFI. To get better insight into the underlying working mechanism of our probe and find the 
locations of the peaks in the QFI, we first briefly revisit a two-level system probe at thermal equilibrium. For a two-level 
system probe of frequency $\omega_{0}$ at thermal equilibrium with the bath of temperature $T$, 
the density matrix of the probe is given by
\begin{equation}
	\hat{\rho}_\text{TLS}=\frac{1}{2}\begin{pmatrix}
	1-\text{tanh}\big(\frac{\omega_0}{2T}\big) & 0\\
	 0 & 1+\text{tanh}\big(\frac{\omega_0}{2T}\big)
	\end{pmatrix}.
\end{equation}\label{eq:DensityTLS}
The QFI associated with this two-level system probe can be evaluated using 
Eq.~(\ref{eq:TLSfisher}), which yields
\begin{equation}
	\mathcal{F}_\text{TLS} = \frac{\omega^2_0\text{sech}^2\big(\frac{\omega_0}{2T}\big)}{4T^4}.
\end{equation}
This expression indicates that the QFI associated with a two-level system probe at thermal equilibrium has a single peak. 
Accordingly, a two-level system with a given energy gap $\omega_{0}$, can only be 
an optimal probe for a single temperature~\cite{campbell2018precision}. At this temperature, $T^{*}$, QFI has a maximum value, 
and the temperature $T^{*}$ can be found by~\cite{campbell2018precision}
\begin{equation}\label{eq:gamma}
	T^{*} = \gamma \omega_{0}, \quad \text{here} \quad 2\gamma 
	= \text{tanh}\bigg(\frac{1}{\gamma}\bigg).
\end{equation}

The diagonal elements of our probe qubit density matrix appear like the two-level 
system probe at thermal equilibrium, except for an additional factor of $\cos\theta_1$. This should not 
be surprising, as the nonlocal dissipation channels in the master equation~(\ref{eq:master}) provide indirect 
access of the thermal bath to our probe. Accordingly, the information about the temperature $T$ of the bath is partially imprinted 
on the diagonal elements of our probe qubit. This is witnessed by the emergence of a peak in the QFI associated with higher 
bath temperature $T^{*}_p$ [see the right peak in Fig.~\ref{fig2}(a)]. The coherent interaction between 
the probe and ancilla provides an additional channel of indirect energy exchange between the bath and the probe. 
This information is imprinted on the off-diagonal elements of our probe's density matrix, which is responsible 
for the emergence of an additional peak at lower temperature $T^*_{1}$ in the QFI.	
	
In the weak coupling limit of ancilla and probe qubits $g\ll (\omega_p,\omega_1)$, 
we can replace $\text{sin}\theta_1\approx\theta_1$, and $\text{cos}\theta_1\approx1$ in the master equation~(\ref{eq:master}). 
In this case, we can write a simple approximate expression for the QFI,
\begin{eqnarray}\label{eq:QFIapprox}
	\mathcal{F}_{Q} &\approx & \mathcal{F}_{0} + \mathcal{F}_{1}, \quad\text{where}\\
	\mathcal{F}_{0} &=& \frac{\omega^2_p\text{sech}^2\big(\frac{\omega_p}{2T}\big)}{4T^4}, \quad \mathcal{F}_{1} = \frac{\frac{1}{2}\theta^2_{1}\omega^2_1\text{sech}^2\big(\frac{\omega_1}{2T}\big)}{4T^4}.
\end{eqnarray}

%------------------------- Figure 3 -------------------------%
	\begin{figure}[t!]
	\centering
		\includegraphics[scale=0.64]{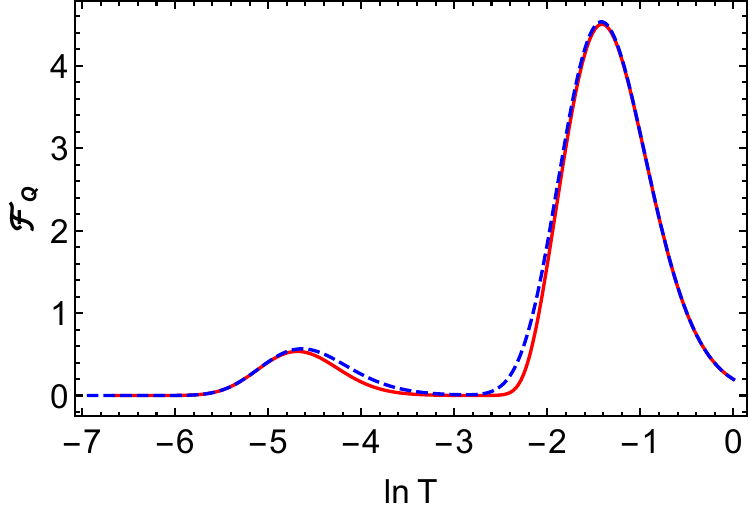}
	\caption{Quantum Fisher information $\mathcal{F}_{Q}$ associated with the qubit probe---considering weak coupling between the ancilla and probe---as a function of temperature $T$ of the thermal bath. The blue-dashed and red-solid lines correspond to approximate and exact QFI, respectively. The approximate expression for the QFI in the weak coupling limit is given by Eq.~(\ref{eq:QFIapprox}), and the exact QFI is evaluated by considering the full expression of QFI, which is too cumbersome to report here. Parameters: $\omega_{p}=1$, $\omega_1=0.04$, and $g=0.01$.}\label{fig3}
	\end{figure}
%-------------------------------------------------------------%

This approximate expression of the QFI can be used to find the location of the peaks in the QFI. 
Note that the two contributions $\mathcal{F}_0$ and $\mathcal{F}_1$ in Eq.~(\ref{eq:QFIapprox}) 
are related to the populations and coherences in the probe.
The temperatures associated with the left and right peaks in Fig.~\ref{fig2}(a) are given by 
$T^*_{1} = \gamma \omega_{1}$, and $T^*_{p} = \gamma \omega_{p}$, respectively.

In Fig.~\ref{fig3}, we depict the approximate and exact QFI given in Eq.~(\ref{eq:QFIapprox}), and Fig.~\ref{fig2}(a), respectively. 
It is evident that our approximate results for QFI overlap with the exact QFI within the weak coupling limit. 
The location of the lower peak can be controlled by the ancilla qubit energy gap $\omega_1$, and baths at lower 
temperatures can be probed with increased accuracy by making the energy gap smaller. In addition, the coupling strength 
between the ancilla and probe qubits can be used to enhance the amplitude of the lower peak, thus enhancing the thermal sensitivity of the probe. 
The right peak in Fig.~\ref{fig3} is identical to the QFI for a two-level system probe at thermal equilibrium.
		
It is important to note that for a single ancilla qubit, the QFI can be approximately given by the sum of two 
$\text{sech}^2\theta$ functions with different amplitudes and centers (Eq.~(\ref{eq:QFIapprox})). 
As a result, two peaks emerge in the QFI, each of which can be identified by the frequencies of the ancilla and probe qubits. 
This hints that including additional ancilla qubits may increase the number of peaks in the QFI. Similar results are reported in a 
periodically driven probe~\cite{Mukherjee2019}; however, the physical mechanism of multiple peaks in the QFI is 
entirely different in these non-autonomous schemes. In the next section, we will investigate the possibility 
of increasing the optimal sensing points of the probe by increasing the number of ancilla qubits.

%-------------------------------------------------------------------------------------------
\subsection{Multiple peaks in QFI with 2 ancilla qubits ($N=2$)}\label{secN2}
%-------------------------------------------------------------------------------------------

To reveal the full advantage of our scheme for the enhancement of low-temperature thermal sensitivity, 
here we consider two ancilla qubits. In this case, the diagonalized Hamiltonian of the ancilla-probe system is given by
\begin{equation}
	\tilde{H} = \frac{1}{2}\sum^2_{k=1}\omega_{k}\tilde{\sigma}^{z}_{k} + \frac{1}{2}\tilde{\Omega}\tilde{\sigma}^{z}_{p},
\end{equation}
where the transformed Pauli matrices are given in Eq.~(\ref{eq:Paulitrans}). The transformed frequency of 
the probe qubit is given by
\begin{eqnarray}\label{eq:omega2}
	\tilde{\Omega} = \Omega_{+} + \Omega_{-}\quad \text{where}\quad  \Omega_{\pm} &=& 
	\sqrt{\omega^2_p\pm4(g_1+g_2)^2}.\nonumber\\
\end{eqnarray}
The master equation for this case remains the same as given in Eq.~(\ref{eq:master}), only the mixing 
angle $\theta_{k}$ is changed to 
\begin{eqnarray}\label{eq:angle2}
\theta_1&=&\frac{1}{2}[\arctan(\frac{2(g_1+g_2)}{\omega_p})+
\arctan(\frac{2(g_1-g_2)}{\omega_p})],\nonumber\\
\theta_2&=&\frac{1}{2}[\arctan(\frac{2(g_1+g_2)}{\omega_p})-
\arctan(\frac{2(g_1-g_2)}{\omega_p})].
\end{eqnarray}
%------------------- Figure 4 ---------------------------%
\begin{figure}[t!]
	\centering
		\includegraphics[scale=0.5]{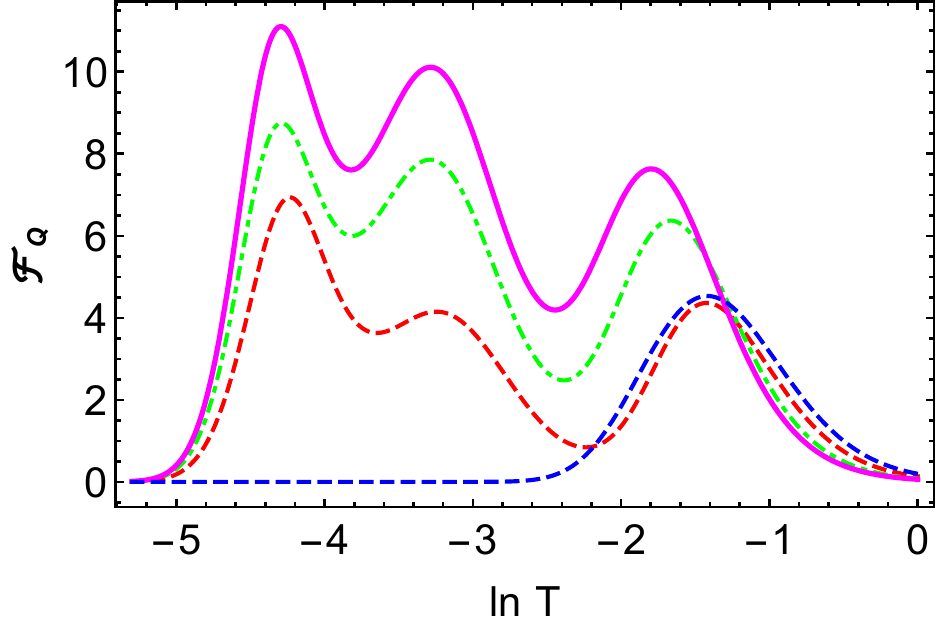}
	\caption{Quantum Fisher information $\mathcal{F}_{Q}$  with respect to the bath temperature 
		$T$ in the case of two ancilla qubits. The solid, dot-dashed, and dashed lines are for 
		$\omega_p=0.26$, $\omega_\text{p}=0.3$, and $\omega_\text{p}=0.4$, respectively. 
		We have added the QFI (blue dashed line) of a two-level system at thermal equilibrium for comparison.
	The rest of the parameters are: $\omega_{1}=0.09$, $\omega_{2}=0.17$, $g_1=0.003$, and $g_2=0.05$.} 
\label{fig4}
\end{figure}
%---------------------------------------------------------%
At the steady state, the ancilla-probe system reaches an equilibrium state under the dissipative dynamics 
governed by the master equation~(\ref{eq:master}). We are only interested in the reduced state of the probe, 
as we wish to perform local measurements on the probe qubit. The state of the probe in a local basis is given 
in Appendix~\ref{App:2Ancilla}, which shows the presence of temperature-dependent coherence induced by both ancilla 
qubits. By the judicious choice of system parameters, QFI exhibits three peaks, as shown in Fig.~\ref{fig4}. 
The emergence of the third peak is associated with additional ancilla qubits in this case. The addition of further 
ancilla qubits will enhance the number of peaks in the QFI, hence enhancing the accuracy of temperature estimation 
and broadening the range of thermal sensitivity.

%-------------------------------------------------------------------------------------------
\section{Global Thermalization Scenario}\label{sec global}
%-------------------------------------------------------------------------------------------
So far, we have employed a scheme to estimate the temperature of the sample through local measurements performed on 
the nonthermal probe qubit, which has a frequency of $\omega_p$ and, has no direct access to the sample.
What happens if the probe qubit also has access to the sample like ancilla qubits? To answer this question, 
we consider a case where all the qubits including the probe are directly coupled to the sample. In this context, 
we assume that all qubits are coupled with the sample and that the interaction between the sample and qubits is weak. 
This interaction is weak enough to allow the total system (composed of the qubits) to reach a global Gibbs state. Since the probe and the ancilla qubits are considered together, they are described by a Gibbs thermal state (GTS) which is given by
\begin{equation}
    \hat{\rho}_\text{GTS}=\frac{\exp(-\beta \hat{H})}{Tr(\exp(-\beta \hat{H}))}.
\end{equation}
We calculate the QFI of the probe using Eq.~(\ref{eq:TLSfisher}) by tracing out the $N$ ancilla 
qubits and considering the reduced state of one qubit (probe).
We first consider the case of two qubits coupled to the sample and trace out one of them. 
Therefore, the reduced state of the probe is given by
\begin{equation}
    \rho_p=Tr_1[\rho_\text{GTS}]=\begin{pmatrix}
    \frac{1}{2}(1-\chi^\prime) & c^\prime \\
    c^\prime & \frac{1}{2}(1+\chi^\prime)
    \end{pmatrix}\label{pobeDglobal},
\end{equation}
where the diagonal and off-diagonal terms are respectively given as
\begin{eqnarray}
\chi^\prime &=&\Omega^\prime\tanh(\frac{\Omega^\prime}{2 T}), \quad\text{and} \\
c^\prime &=& \frac{g}{\Omega^\prime}\tanh(\frac{\Omega^\prime}{2T})\tanh (\frac{\omega_1}{2 T}),
\label{eq:cohG}
\end{eqnarray}
where $\Omega^\prime=\sqrt{\omega_p^2+4g^2}$. The density matrices of the probe [see Eq.~(\ref{ProbD}) and 
Eq.~(\ref{pobeDglobal})] are quite similar for both the local and global thermalization cases.
The full expression for QFI is too cumbersome again, however, if we do not consider an ultrastrong 
coupling regime
and remain within weak regime $\omega_p\gg\omega_1\geq g$, the QFI for the qubit probe~\cite{PhysRevA.87.022337} 
can be written as [see Eq.~(\ref{eq:TLSfisher})]
\begin{eqnarray}\label{eq:QFIapproxG}
	\mathcal{F}_{Q} &\approx & \mathcal{F}_{low} + \mathcal{F}_{high},\quad\text{where}\\
	\mathcal{F}_{low} &=& \frac{8g^2\omega^2_1\sinh^6\big(\frac{\omega_1}{2T}\big)}{T^4\sinh^4(\frac{\omega_1}{2T})}, \\
		\mathcal{F}_{high} &=& \frac{\omega^2_p \text{sech}^2(\frac{\omega_p}{2T})}{4T^4(1+2g^2(1+\cosh(\frac{\omega_p}{2T})))}.
\end{eqnarray}
These approximate expressions of QFI are related to the populations and coherences of the probe and denote the peaks for lower and higher temperatures in Fig.~\ref{fig5}(a), respectively.
Figure~\ref{fig5}(a) depicts both the approximate and exact QFI, and it is evident that they agree well within the considered parameter range. The QFI shows two peaks when there is a single ancilla qubit. 
In the limit of $g\rightarrow 0 $, the QFI reduces to the Fisher information of a single qubit in a thermal state, as expected. 
The approximate expression for the QFI enables us to determine the optimal temperature $T^*_\text{lower}$ 
for the lower peak, where the QFI is maximum. Specifically, we find that $T^*_\text{lower} \approx \omega_1/4$. In the subsequent two sections, we discuss the results for the case of thermalized identical and non-identical qubits.

%-------------------------------------------------------------------------------------------
\subsection{Identical qubits}\label{sec ident}
\begin{figure}[t]
		\centering
		\subfloat[]{
			\includegraphics[scale=0.161]{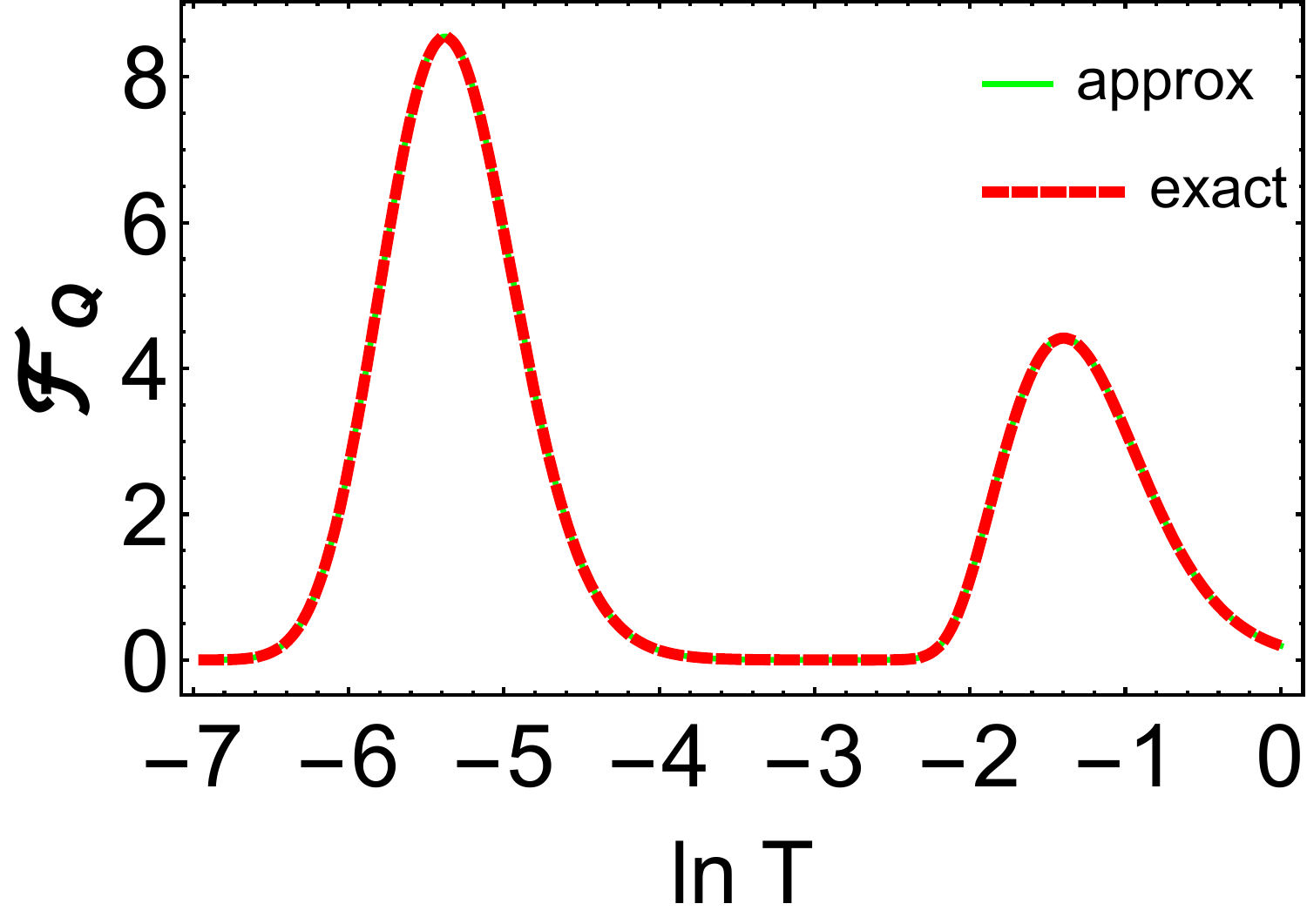}}
		\subfloat[]{
			\includegraphics[scale=0.1725]{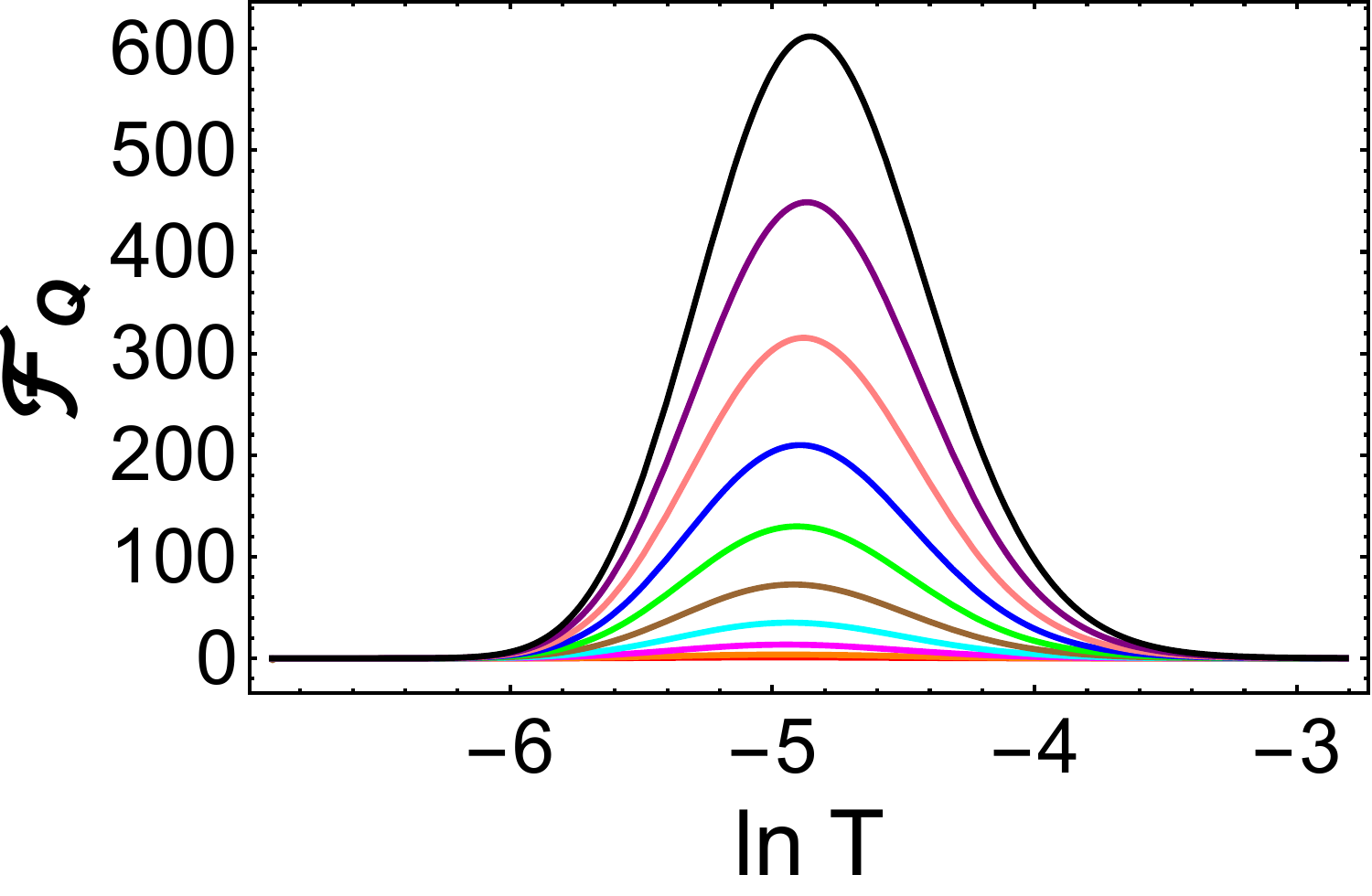}}\\
			\subfloat[]{
			\includegraphics[scale=0.167]{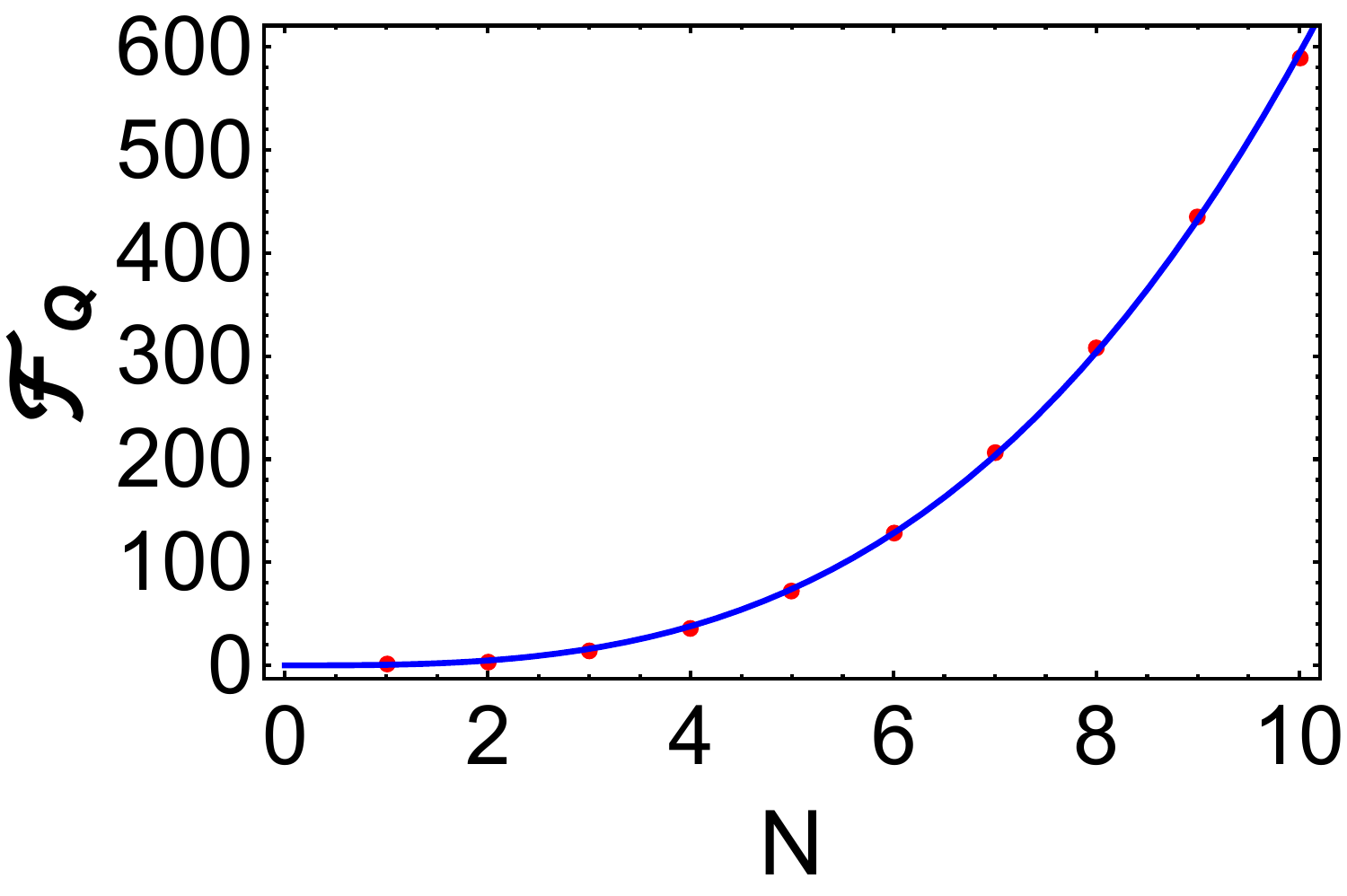}}
			\subfloat[]{
			\includegraphics[scale=0.167]{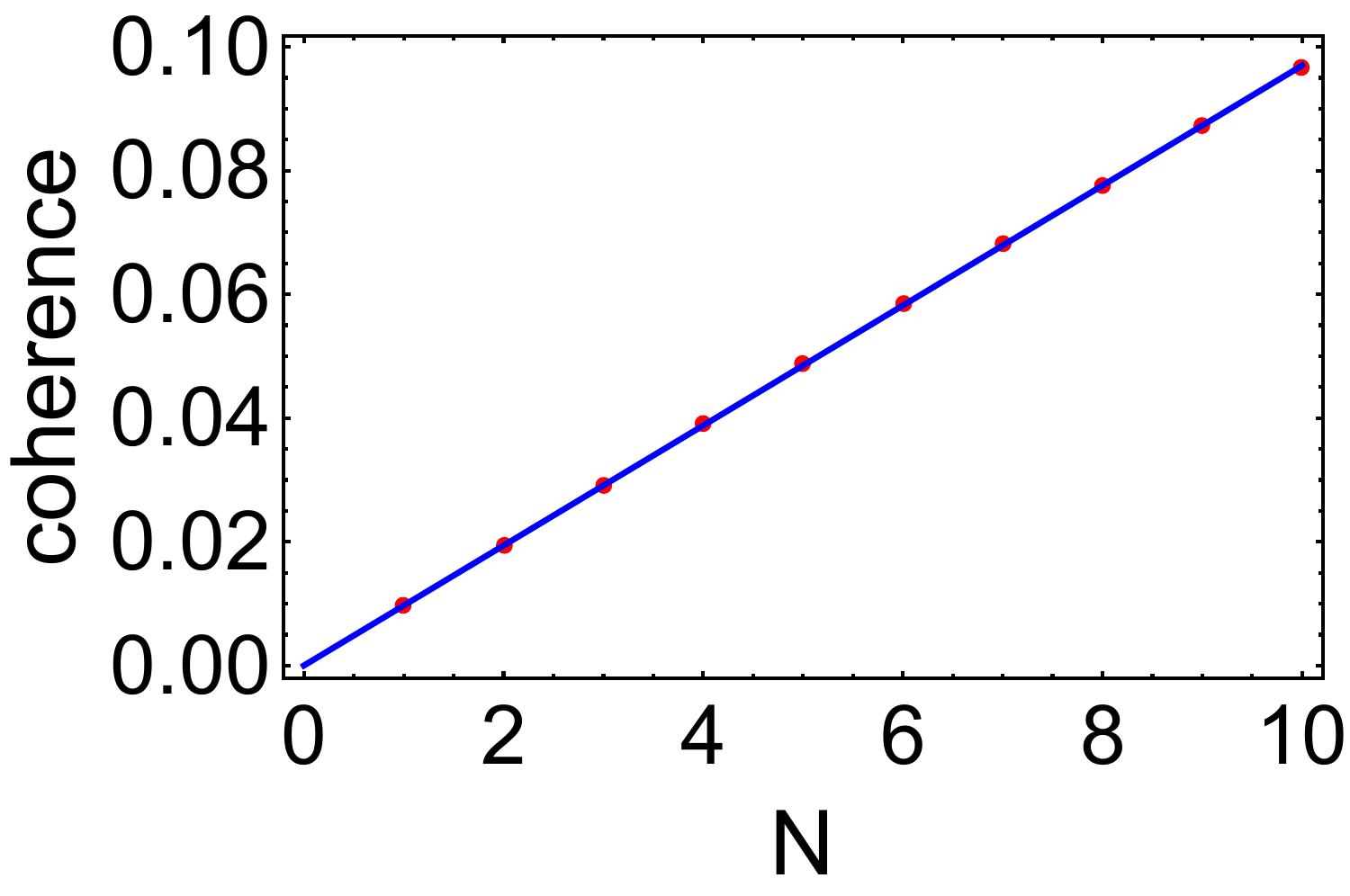}}
		\caption{\textbf{(a)} Quantum Fisher information $\mathcal{F}_{Q}$ as a function of temperature $T$ for a single ancilla qubit. The green and
red-dashed lines represent the approximate and exact Fisher information. The parameters are: $\omega_{p}=1$, $\omega_{1}=0.02$, $g=0.02$. 
\textbf{(b)} The QFI with respect to $T$ for different number of qubits. The QFI increases as $N$ is increased, which is denoted by the curves 
from bottom to top with increasing $N$, the parameters are $\omega_p=1$, $\omega=0.03$, $g=0.01$, and $N=10$. \textbf{(c)},\textbf{(d)} 
The QFI and coherence for low temperatures with respect to the number of qubits attached to the sample. The blue curve corresponds to the 
QFI and the corresponding scaling is represented by red dotted points. The parameters chosen here are as follows: $\omega_p=1$, $\omega=0.03$, 
and $g=0.01$.}\label{fig5}
	\end{figure}

%-------------------------------------------------------------------------------------------
We study the effect of many qubits on the information flow from the sample and observe how the number of qubits affects the 
QFI. To this end, we assume that all the qubits are identical such that $\omega_k=\omega_p=\omega$ and $g_k=g$ and observe 
how the behavior of QFI changes as one increases the number of qubits for low-$T$ sensitivity. This effect is presented in 
Fig.~\ref{fig5}(b), where the peak of QFI at low-temperature increases with $N$. To better understand this phenomenon, we 
plotted QFI as a function of the number of qubits, $N$, in Fig.~\ref{fig5}(c), where the QFI of the probe increases with $N$ 
as depicted by the cubic behavior. Since one of the features of our model is the coherence generation in the probe state; 
therefore, we observe that the coherence in the probe linearly scales with $N$. 
In the strong coupling regime, coherence does not exhibit a linear behavior and instead saturates at an upper 
limit of 1/2 for a relatively small $N$. The QFI behavior and coherence in this regime are interrelated. 
The QFI is represented by a solid blue line that approximately scales as $aN^3$, where $a\sim 0.5$. For $N=10$, 
coherence increases linearly with $\sim N$, resulting in an increase in QFI that scales as $\sim aN^3$.

%------------------- Figure 5 ---------------------------%
  %---------------------------Figure 5--------------------------%
  
%-------------------------------------------------------------------------------------------
\subsection{Non-identical qubits}\label{non ident}
%-------------------------------------------------------------------------------------------

\begin{figure}[t!]
	\centering
		\includegraphics[scale=0.65]{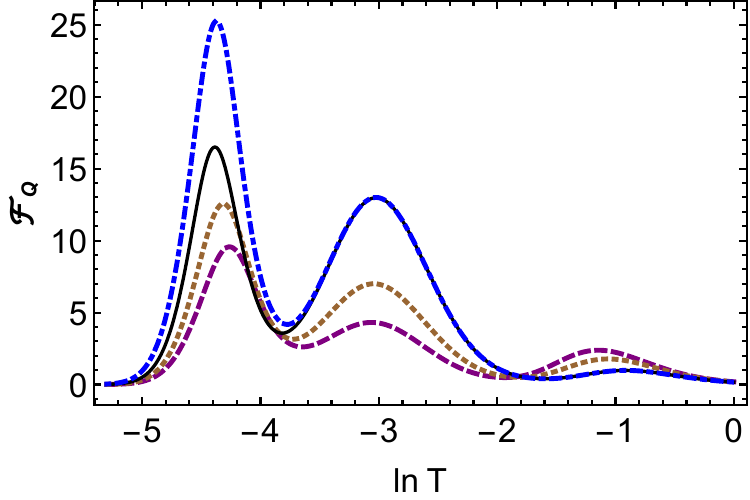}
	\caption{Quantum Fisher information $\mathcal{F}_{Q}$  with respect to the bath temperature $T$ for $N=4$ when the qubits are non-identical. 
		The dashed magenta, the brown dotted, the solid black, and the dot-dashed blue curves correspond to the parameters 
		$g_2 = 0.15$, $g_2 = 0.2$, $g_2 = 0.3$ and, $g_2 = 0.3$ with $\omega_1 = 0.85$. The other parameters are: $\omega_p = 1$, 
		$\omega_1= 0.09$, $\omega_2 = 0.2$, $\omega_3 = 0.5$, $g_1= 0.003$, and $g_3 = 0.008$.} \label{fig6}
\end{figure}

The presence of multiple peaks in QFI can still be detected in scenarios where all qubits connected to the sample are thermalized with the bath, and each qubit possesses a distinct frequency and coupling strength. Figure~\ref{fig6} shows the multiple peaks in QFI for the case of $N = 4$. With the help of proper tuning of the parameters, one can set the desired range for low and high temperatures. In this case, the peak at lower $T$ has more amplitude than in the previous case. Continuing with this same strategy, we can increase the thermal sensitivity at higher temperatures, but we do not report the results here. If we increase the number of qubits, the peaks of QFI become smoother, and the dips begin to disappear, resulting in an enhancement of low-$T$ sensitivity.

%------------------- Figure 6 ---------------------------%
%---------------------------------------------------------%

%-------------------------------------------------------------------------------------------
\section{Conclusion}\label{sec conc}
%-------------------------------------------------------------------------------------------

Our study has demonstrated the potential of quantum coherence to enhance the thermal sensitivity of a two-level system probe and broaden its range of low-temperature estimation. We have proposed a simple and precise scheme where ancilla qubits indirectly couple the probe to a sample in thermal equilibrium. The ancilla qubits generate coherence in the probe, which encodes temperature information on both the diagonal and off-diagonal elements of its density matrix, leading to multiple peaks in the QFI at low temperatures depending on the number of ancilla qubits. Typically, a probe is assumed to be in thermal equilibrium with the sample. In this case, temperature information is encoded only on the diagonal elements of the probe's density matrix. Accordingly, the QFI becomes equal to the classical information and, other than the discreteness of energy level, the quantum probe does not exhibit any quantumness. A two-level system probe at thermal equilibrium can only estimate a single temperature with optimal accuracy, which can be seen by the existence of a single peak in the QFI. On the contrary, we have shown that in the presence of ancilla qubits, nonlocal dissipation channels in the global master equation imprint temperature information on the populations and on the coherence of the probe. Furthermore, our global master equation approach is consistent with the laws of thermodynamics and does not require highly degenerate excited states or energetically costly nonautonomous schemes. Finally, to study the many qubit effects (such as ancilla qubits) on the QFI of the probe, we presented the results when all the qubits, including the probe, have direct access to the sample where the probe qubit is coupled to all the ancilla qubits. We summarized that the QFI of the probe at thermal equilibrium scales as $\sim aN^3$, meaning that the sensitivity at low temperatures increases with the number of ancilla qubits.

Our findings have broad implications for the development of quantum thermometry and can be extended to other physical systems such as multimode optomechanical systems~\cite{RevModPhys.86.1391, Massel2012}, which is another asymmetric interaction, i.e.,  a bosonic version of the one we considered here. The implementation of precision thermometry is also crucial to understanding Fermi gases where ultra-low temperature measurements are significant for the practical realization and applications of degenerate Fermi gases~\cite{Onofrio_2016}.
The ability to precisely measure a broad range of low temperatures is crucial for various quantum technological applications, and our proposed scheme can contribute to advancing these fields.

%------------------------------------------------------------------------------------
\section*{Acknowledgment }
%------------------------------------------------------------------------------------

\"{O}.~E.~M.~acknowledges support from the Scientific and Technological Research Council of T\"urkiye (T\"UBITAK) under Grant No.~122F371.

%------------------------------------------------------------------------------------
\appendix
%------------------------------------------------------------------------------------

%---------------------------------------------------------------------------------------
\section{The state of probe for symmetric ancilla-probe interaction}\label{app:A}
%---------------------------------------------------------------------------------------

Here we consider the dipole-dipole interaction between the probe and ancilla qubit to analyze whether such a symmetric interaction can generate coherence in the probe. The Hamiltonian of the probe-qubit system interacting via dipole-dipole interaction is given by 
\begin{equation}
    \hat{H}_\text{dd}=\frac{1}{2}\omega_p\hat{\sigma}^z_p+\frac{1}{2}\omega_1
    \hat{\sigma}^z_1+g(\hat{\sigma}^+_1\hat{\sigma}^-_p+\hat{\sigma}^-_1\hat{\sigma}^+_p).
\end{equation}
In its diagonal form, the above Hamiltonian can be written as~\cite{Levy_2014}
\begin{equation}\label{eq:dd}
    \tilde{H}_\text{dd}=\frac{\omega_+}{2}\tilde{\sigma}^z_p+\frac{\omega_-}
    {2}\tilde{\sigma}^z_1,
\end{equation}
where the transformed frequencies $\omega_{\pm}$ and the angle $\theta$ are given by
\begin{eqnarray}
    \omega_{\pm}&=&\frac{\omega_1+\omega_p}{2} \pm \sqrt{\bigg(\frac{\omega_1-\omega_p}{2}\bigg)^2+g^2},\\ \cos^2\theta&=&\frac{\omega_1-\omega_p}{\omega_+-\omega_-} \label{eqA3}.
\end{eqnarray}
The transformed Pauli operators in Eq.~(\ref{eq:dd}) have the form
\begin{eqnarray}\label{eq:ddtrans}
\tilde{\sigma}_\alpha^{z} &=& 2\hat{\sigma}_\alpha^+\hat{\sigma}_\alpha^--\mathbb{1}, \quad \tilde{\sigma}^-_p = \hat{\sigma}^-_p\text{cos}\theta + \hat{\sigma}^-_1\text{sin}\theta,\\ \tilde{\sigma}^-_1 &=& \hat{\sigma}^-_1\text{sin}\theta - \hat{\sigma}^-_p\text{cos}\theta.
\end{eqnarray}
Recall that only the ancilla is coupled to the thermal bath in our scheme; in such a setup, the joint state of the ancilla-probe system can be found by writing a global master equation~\cite{Levy_2014}. At the steady state, the joint state of the ancilla-probe system is diagonal because of the imposition of rotating wave approximation in the derivation of the master equation, and it is given by~\cite{Levy_2014}
\begin{equation}
    \Tilde{\rho}_\text{dd}=\Tilde{\rho}_p\otimes\Tilde{\rho}_1,\label{A7}
\end{equation}
where $\Tilde{\rho}_p$ and $\Tilde{\rho}_1$ are the reduced states of the probe  and ancilla qubit in the transformed basis and these are given by
\begin{eqnarray}
  \Tilde{\rho}_p &=&
  \begin{pmatrix}
 \frac{1}{1+e^{\beta\omega_+}} & 0 \\
 0 & 1-\frac{1}{1+e^{\beta\omega_+}} 
\end{pmatrix}, \\
\Tilde{\rho}_1 &=&
  \begin{pmatrix}
 \frac{1}{1+e^{\beta\omega_-}} & 0 \\
 0 & 1-\frac{1}{1+e^{\beta\omega_-}}
\end{pmatrix},\label{A6}
\end{eqnarray}
where the frequencies $\omega_{\pm}$ are given in Eq.~(\ref{eqA3}). We are interested in the state of the probe in a local basis, as we wish to perform local measurements on the probe for temperature estimation. The reduced state of the probe in the local basis can be found by using the transformation given in Eq.~(\ref{eq:ddtrans}) to convert the global master equation into the local master equation. At the steady state, the state of the probe in the local basis is given by~\cite{Levy_2014}
\begin{equation}
  \hat{\rho}_p=
  \begin{pmatrix}
 \frac{1}{2} & 0 \\
 0 & \frac{1}{2} 
\end{pmatrix},
\end{equation}
which is a maximally mixed state. Therefore, it is evident that, in our scheme, the dipole-dipole interaction does not generate coherences in the state of probe qubit.

%---------------------------------------------------------------------------------------
\section{The state of probe for anti-symmetric ancilla-probe interaction}\label{app:B}
%---------------------------------------------------------------------------------------

 In Appendix.~\ref{app:A}, we showed that it is not possible to generate coherences in the probe qubit by considering symmetric interaction like dipole-dipole interaction. Here we consider the anti-symmetric interaction between the ancilla and probe qubits and investigate the possibility of coherence generation in the probe qubit. To this end, we consider Dzyaloshinskii-Moriya (DM) interaction between the ancilla and probe qubits. For simplicity, we assume that the anisotropy field is aligned in the $\textit{z}$ direction, so that the Hamiltonian of the system is given by~\cite{PhysRevE.104.054137}
\begin{equation}
    \hat{H}_\text{DM}=\frac{1}{2}\omega_{1}\hat{\sigma}^z_{1}+\frac{1}{2}\omega_{p}\hat{\sigma}^z_{p} + g(\hat{\sigma}^x_1\hat{\sigma}^y_p-\hat{\sigma}^y_1\hat{\sigma}^x_p).
\end{equation}
The Hamiltonian of two qubits with such interaction can be transformed into its diagonal basis with the help of the following transformation,
\begin{equation}\label{eq:DMtrans}
\hat{U}=\cos^2(\theta/2)+\sin^2(\theta/2)\hat{\sigma}^z_1\hat{\sigma}^z_p+i\frac{\sin\theta}{2}(\hat{\sigma}^x_1\hat{\sigma}^x_p+\hat{\sigma}^y_1\hat{\sigma}^y_p).
\end{equation}
and the diagonalized Hamiltonian for two qubits coupled via DM interaction has the following form
\begin{equation}
    \tilde{H}_{DM}=\frac{(\omega_S+\Omega)}{2}\tilde{\sigma}^z_p+\frac{(\omega_S-\Omega)}{2}\tilde{\sigma}^z_1,
\end{equation}
where
\begin{equation}
    \omega_{S}:=\frac{\omega_1+\omega_p}{2}, \quad \Omega:=\sqrt{\omega_D+4g^2}\label{B4}
\end{equation}
with $\omega_D:=(\omega_1-\omega_p)/2$ introduced as a notation for the sake of convenience. 
In the local basis, the eigenvectors associated with the eigenvalues are expressed as
	\begin{eqnarray}\label{eq:eigenstate1}
	\ket{1} &=& \ket{++},\\ 	
	\ket{2} &=& \cos\theta\ket{+-}+i\sin\theta\ket{-+}, \\
	\ket{3} &=&  \cos\theta\ket{-+}+i\sin\theta\ket{+-},\\ 
	\ket{4} &=& \cos\theta\ket{--} 
\label{eq:B5},
	\end{eqnarray}
where the angle $\theta$ is defined by

%------------------------- Relative Error bound -------------------------%
	\begin{figure}[b!]
	\centering
		\includegraphics[scale=0.62]{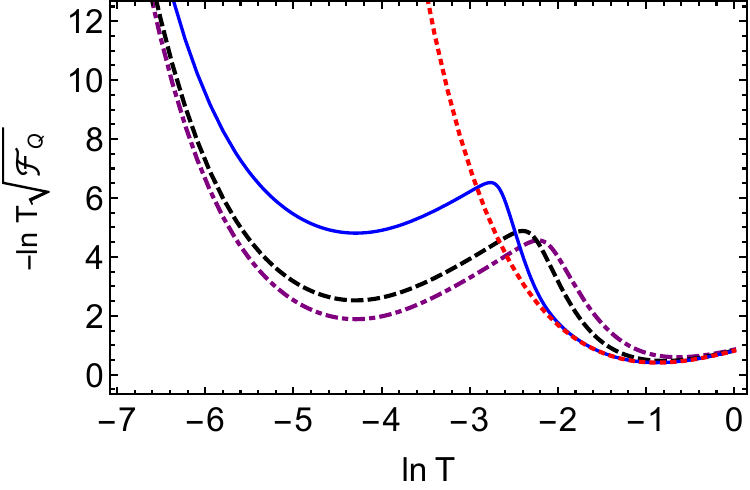}
	\caption{The relative error bound $\delta T/T=1/(T\sqrt{\mathcal{F}_Q})$ for the estimation of bath temperature $T$ for different coupling strengths between the probe and ancilla qubit. The dot-dashed, dashed and solid curves correspond to $g=0.2,0.1,0.01$, respectively. For comparison, we superimposed the red dotted curve for a two-level system at thermal equilibrium with the bath of temperature $T$. The other parameters are $\omega_{p}=1$ and $\omega_1=0.04$.}\label{error}
	\end{figure}
%-------------------------------------------------------------%
	
\begin{eqnarray}
    \cos\theta &=& \frac{2g}{\sqrt{4g^2+(\omega_D-\Omega)^2}}, \\ 
    \sin\theta &=& \frac{\omega_D-\Omega}{\sqrt{4g^2+(\omega_D-\Omega)^2}}
\end{eqnarray}
The global master equation for two qubits coupled via DM interaction can be found in Ref.~\cite{PhysRevE.104.054137}. In the basis in which Hamiltonian is diagonal, the joint state of the ancilla and probe qubits at the steady state is similar to Eq.~(\ref{A7}), with a difference that the frequencies $\omega_{+}$ and $\omega_{-}$ are replaced by $\omega_{s}$ and $\Omega$, respectively. The joint state of ancilla-probe qubits can be transformed back to the local basis using the transformation in Eq.~(\ref{eq:DMtrans}). At the steady state, the reduced state of the probe in the local basis is given by~\cite{PhysRevE.104.054137}
\begin{widetext}
\begin{equation}
  \hat{\rho}_p=\frac{1}{\mathcal{K}}
  \begin{pmatrix}
 e^{\Omega/T}+e^{\omega_s/T}(e^{2\Omega/T}\text{cos}^2\theta+\text{sin}^2\theta) & 0 \\
 0 & e^{\omega_s/T}(e^{(\omega_s+\Omega)/T}+\text{cos}^2\theta+e^{2\Omega/T}\text{sin}^2\theta)
\end{pmatrix},
\end{equation}
\end{widetext}
where we define $\mathcal{K}$ as
\begin{equation}
\mathcal{K} = (e^{\Omega/T}+e^{\omega_s/T})(1+e^{(\omega_s+\Omega)/T}).
\end{equation}
The reduced density matrix of the probe shows that anti-symmetric DM interaction does not generate coherence; as a result, it can not enhance the thermal sensitivity of the probe if employed in our scheme.

%-----------------------------------------------------------------------
\section{Relative error bound for probe qubit}\label{app:relative}
%-----------------------------------------------------------------------

In Fig.~\ref{error}, we plot the relative error bound $\delta T/T=1/(T\sqrt{\mathcal{F}_Q})$ for the probe qubit as a function of bath temperature $T$ for different values of coupling strength $g$. It is immediately evident from the results that the range of thermal sensitivity is enhanced by considering our scheme based on asymmetric interaction between the probe and ancilla qubits. In contrast, the relative error bound for a two-level system probe at thermal equilibrium (red dotted curve in Fig.~\ref{error}) diverges at much higher temperatures. Hence we conclude that implanting our thermometry scheme based on asymmetric interaction between the probe and ancilla qubits can significantly boost the probe's thermal sensitivity at lower temperatures. Unlike QFI, no additional peak can be observed in this quantity since this feature is not general for any number of qubits, $N$.

%-----------------------------------------------------------------------
\section{ The reduced state of probe qubit for N=2}\label{App:2Ancilla}
%-----------------------------------------------------------------------
	
If the probe qubit is coupled to two ancilla qubits, then the analytical expression of the reduced density matrix for the probe qubit can be calculated from the steady state of the joint ancilla-probe density matrix. At the steady state, the joint state of the ancilla-probe qubits in the global basis is given by
\begin{equation}
\tilde{\rho}_\text{ss}=\tilde{\rho}_1\otimes\hat{\rho}_2\otimes\tilde{\rho}_p.
\end{equation}
The reduced density matrix of the probe qubit in the local basis can be obtained by first writing the joint density matrix in the local basis using the transformation given in Eq.~(\ref{eq:transformation}). Then taking a partial trace over the ancilla qubits, the resulting reduced state of the probe qubit is given by
\begin{equation}
\hat{\rho}_p=\text{Tr}_{1,2}[\tilde{\rho}_\text{ss}],
\end{equation}
and
\begin{equation}
\hat{\rho}_p=
\begin{pmatrix}
\frac{1}{2}(1-\alpha\beta) & c^\prime\\
c^\prime& \frac{1}{2}(1+\alpha\beta)
\end{pmatrix},
\end{equation}
where $\alpha$, $\beta$, and $c^\prime$ are given by
\begin{equation}
\alpha=\tanh(\frac{\tilde{\Omega}}{2 T}),
\end{equation}
\begin{equation}
\beta=\Big[\cos(\theta_1)\cos(\theta_2)-\sin(\theta_1)\sin(\theta_2)\tanh(\frac{\omega_1}{2T})\tanh(\frac{\omega_2}{2T})\Big],
\end{equation}
\begin{widetext}
\begin{equation}
c^\prime=\frac{(e^{\frac{\tilde{\Omega}}{T}}-1)\Big[\sin (\theta_1-\theta_2)(e^{\omega_1/T}-e^{\omega_2/T})+\sin (\theta_1+\theta_2) (e^{\frac{\omega_1+\omega_2}{T}}-1)\Big]}{2 (e^{\omega_1/T}+1) (e^{\omega_2/T}+1)(e^{\frac{\tilde{\Omega}}{T}}+1)}.
\end{equation}	
\end{widetext}
	
%-----------------------------------------------------------------------
\section{Experimental feasibility}\label{exp}
%-----------------------------------------------------------------------

Our model of the many-qubit system is based on asymmetric interaction, which is responsible for the generation of coherence in the probe. In terms of Hamiltonian implementations [i.e., Eq.~(\ref{eq1})], we provide some examples of realization and possible paths for such interactions.

(i) The asymmetric interaction given in Eq.~(\ref{eq1}) can be realized in a system comprised of a mechanical resonator coupled to an optical resonator~\cite{PhysRevA.49.433,PhysRevA.51.2537}. By applying the Holstein–Primakoff transformation to the bosons and phonons under the assumptions of weakly excited spins, this bosonic optomechanical model can be mapped to the asymmetric spin-spin coupling model, where only the two lowest vibronic levels are accessible~\cite{Moqadam2015}. 

(ii) The coupled Raman model, consisting of a three-level atom in a single-mode cavity, is another possible route for such interactions~\cite{Phoenix1990}. To obtain the desired asymmetric qubit-qubit interaction, assume the weak excitation of the cavity mode and replace the bosonic operators ($\hat{a}^\dagger\hat{a}$) with spin operators $\hat{\sigma}^z$.

(iii) Another suitable scheme is the circuit QED, (see Ref.~\cite{PhysRevA.69.062320}), where the junction parameters are tuned to obtain the phase-gate term described in Ref.~\cite{PhysRevLett.123.180602}. This term takes the form of $\hat{\sigma}^z\hat{\sigma}^x$ when the resonator is only slightly excited. 

(iv) Trapped ions can be utilized to simulate the quantum walk on a circle, whereby the walker's movements are carried out in the quantum optical phase space using a single step generator ($U=\exp{(ip\sigma^zH)}$)~\cite{PhysRevA.65.032310}. If the vibrational excitation is significantly small, the step generator's effective Hamiltonian corresponds to an asymmetric spin-spin interaction. 

(v) Based on the parameter range we considered, another possible example can be the spin-boson model, which has been studied extensively for metrological purposes. Under weak excitation conditions, the spin-boson model's interaction term can also be considered as an asymmetric interaction.

%-----------------------------------------------------------------------
%\nocite{*}
%\bibliographystyle{apsrev4-2.bst}
\bibliography{qtm}

%apsrev4-2.bst 2019-01-14 (MD) hand-edited version of apsrev4-1.bst
%Control: key (0)
%Control: author (8) initials jnrlst
%Control: editor formatted (1) identically to author
%Control: production of article title (0) allowed
%Control: page (0) single
%Control: year (1) truncated
%Control: production of eprint (0) enabled
\begin{thebibliography}{79}%
\makeatletter
\providecommand \@ifxundefined [1]{%
 \@ifx{#1\undefined}
}%
\providecommand \@ifnum [1]{%
 \ifnum #1\expandafter \@firstoftwo
 \else \expandafter \@secondoftwo
 \fi
}%
\providecommand \@ifx [1]{%
 \ifx #1\expandafter \@firstoftwo
 \else \expandafter \@secondoftwo
 \fi
}%
\providecommand \natexlab [1]{#1}%
\providecommand \enquote  [1]{``#1''}%
\providecommand \bibnamefont  [1]{#1}%
\providecommand \bibfnamefont [1]{#1}%
\providecommand \citenamefont [1]{#1}%
\providecommand \href@noop [0]{\@secondoftwo}%
\providecommand \href [0]{\begingroup \@sanitize@url \@href}%
\providecommand \@href[1]{\@@startlink{#1}\@@href}%
\providecommand \@@href[1]{\endgroup#1\@@endlink}%
\providecommand \@sanitize@url [0]{\catcode `\\12\catcode `\$12\catcode
  `\&12\catcode `\#12\catcode `\^12\catcode `\_12\catcode `\%12\relax}%
\providecommand \@@startlink[1]{}%
\providecommand \@@endlink[0]{}%
\providecommand \url  [0]{\begingroup\@sanitize@url \@url }%
\providecommand \@url [1]{\endgroup\@href {#1}{\urlprefix }}%
\providecommand \urlprefix  [0]{URL }%
\providecommand \Eprint [0]{\href }%
\providecommand \doibase [0]{https://doi.org/}%
\providecommand \selectlanguage [0]{\@gobble}%
\providecommand \bibinfo  [0]{\@secondoftwo}%
\providecommand \bibfield  [0]{\@secondoftwo}%
\providecommand \translation [1]{[#1]}%
\providecommand \BibitemOpen [0]{}%
\providecommand \bibitemStop [0]{}%
\providecommand \bibitemNoStop [0]{.\EOS\space}%
\providecommand \EOS [0]{\spacefactor3000\relax}%
\providecommand \BibitemShut  [1]{\csname bibitem#1\endcsname}%
\let\auto@bib@innerbib\@empty
%</preamble>
\bibitem [{\citenamefont {Mehboudi}\ \emph
  {et~al.}(2019{\natexlab{a}})\citenamefont {Mehboudi}, \citenamefont
  {Sanpera},\ and\ \citenamefont {Correa}}]{Mehboudi_2019}%
  \BibitemOpen
  \bibfield  {author} {\bibinfo {author} {\bibfnamefont {M.}~\bibnamefont
  {Mehboudi}}, \bibinfo {author} {\bibfnamefont {A.}~\bibnamefont {Sanpera}},\
  and\ \bibinfo {author} {\bibfnamefont {L.~A.}\ \bibnamefont {Correa}},\
  }\bibfield  {title} {\bibinfo {title} {Thermometry in the quantum regime:
  Recent theoretical progress},\ }\href
  {https://doi.org/10.1088/1751-8121/ab2828} {\bibfield  {journal} {\bibinfo
  {journal} {J. Phys. A: Math. Theor.}\ }\textbf {\bibinfo {volume} {52}},\
  \bibinfo {pages} {303001} (\bibinfo {year} {2019}{\natexlab{a}})}\BibitemShut
  {NoStop}%
\bibitem [{\citenamefont {Giovannetti}\ \emph {et~al.}(2011)\citenamefont
  {Giovannetti}, \citenamefont {Lloyd},\ and\ \citenamefont
  {Maccone}}]{giovannetti2011advances}%
  \BibitemOpen
  \bibfield  {author} {\bibinfo {author} {\bibfnamefont {V.}~\bibnamefont
  {Giovannetti}}, \bibinfo {author} {\bibfnamefont {S.}~\bibnamefont {Lloyd}},\
  and\ \bibinfo {author} {\bibfnamefont {L.}~\bibnamefont {Maccone}},\
  }\bibfield  {title} {\bibinfo {title} {Advances in quantum metrology},\
  }\href {https://doi.org/10.1038/nphoton.2011.35} {\bibfield  {journal}
  {\bibinfo  {journal} {Nat. Photonics}\ }\textbf {\bibinfo {volume} {5}},\
  \bibinfo {pages} {222} (\bibinfo {year} {2011})}\BibitemShut {NoStop}%
\bibitem [{\citenamefont {Giovannetti}\ \emph {et~al.}(2006)\citenamefont
  {Giovannetti}, \citenamefont {Lloyd},\ and\ \citenamefont
  {Maccone}}]{PhysRevLett.96.010401}%
  \BibitemOpen
  \bibfield  {author} {\bibinfo {author} {\bibfnamefont {V.}~\bibnamefont
  {Giovannetti}}, \bibinfo {author} {\bibfnamefont {S.}~\bibnamefont {Lloyd}},\
  and\ \bibinfo {author} {\bibfnamefont {L.}~\bibnamefont {Maccone}},\
  }\bibfield  {title} {\bibinfo {title} {Quantum metrology},\ }\href
  {https://doi.org/10.1103/PhysRevLett.96.010401} {\bibfield  {journal}
  {\bibinfo  {journal} {Phys. Rev. Lett.}\ }\textbf {\bibinfo {volume} {96}},\
  \bibinfo {pages} {010401} (\bibinfo {year} {2006})}\BibitemShut {NoStop}%
\bibitem [{\citenamefont {T{\'{o}}th}\ and\ \citenamefont
  {Apellaniz}(2014)}]{toth2014quantum}%
  \BibitemOpen
  \bibfield  {author} {\bibinfo {author} {\bibfnamefont {G.}~\bibnamefont
  {T{\'{o}}th}}\ and\ \bibinfo {author} {\bibfnamefont {I.}~\bibnamefont
  {Apellaniz}},\ }\bibfield  {title} {\bibinfo {title} {Quantum metrology from
  a quantum information science perspective},\ }\href
  {https://doi.org/10.1088/1751-8113/47/42/424006} {\bibfield  {journal}
  {\bibinfo  {journal} {J. Phys. A: Math. Theor.}\ }\textbf {\bibinfo {volume}
  {47}},\ \bibinfo {pages} {424006} (\bibinfo {year} {2014})}\BibitemShut
  {NoStop}%
\bibitem [{\citenamefont {Celi}\ \emph {et~al.}(2017)\citenamefont {Celi},
  \citenamefont {Sanpera}, \citenamefont {Ahufinger},\ and\ \citenamefont
  {Lewenstein}}]{celi2016quantum}%
  \BibitemOpen
  \bibfield  {author} {\bibinfo {author} {\bibfnamefont {A.}~\bibnamefont
  {Celi}}, \bibinfo {author} {\bibfnamefont {A.}~\bibnamefont {Sanpera}},
  \bibinfo {author} {\bibfnamefont {V.}~\bibnamefont {Ahufinger}},\ and\
  \bibinfo {author} {\bibfnamefont {M.}~\bibnamefont {Lewenstein}},\ }\bibfield
   {title} {\bibinfo {title} {Quantum optics and frontiers of physics: The
  third quantum revolution},\ }\href
  {https://doi.org/10.1088/1402-4896/92/1/013003} {\bibfield  {journal}
  {\bibinfo  {journal} {Phys. Scr.}\ }\textbf {\bibinfo {volume} {92}},\
  \bibinfo {pages} {013003} (\bibinfo {year} {2017})}\BibitemShut {NoStop}%
\bibitem [{\citenamefont {Bloch}\ \emph {et~al.}(2008)\citenamefont {Bloch},
  \citenamefont {Dalibard},\ and\ \citenamefont {Zwerger}}]{RevModPhys.80.885}%
  \BibitemOpen
  \bibfield  {author} {\bibinfo {author} {\bibfnamefont {I.}~\bibnamefont
  {Bloch}}, \bibinfo {author} {\bibfnamefont {J.}~\bibnamefont {Dalibard}},\
  and\ \bibinfo {author} {\bibfnamefont {W.}~\bibnamefont {Zwerger}},\
  }\bibfield  {title} {\bibinfo {title} {Many-body physics with ultracold
  gases},\ }\href {https://doi.org/10.1103/RevModPhys.80.885} {\bibfield
  {journal} {\bibinfo  {journal} {Rev. Mod. Phys.}\ }\textbf {\bibinfo {volume}
  {80}},\ \bibinfo {pages} {885} (\bibinfo {year} {2008})}\BibitemShut
  {NoStop}%
\bibitem [{\citenamefont {Bloch}\ \emph {et~al.}(2012)\citenamefont {Bloch},
  \citenamefont {Dalibard},\ and\ \citenamefont
  {Nascimb{\`e}ne}}]{bloch2012quantum}%
  \BibitemOpen
  \bibfield  {author} {\bibinfo {author} {\bibfnamefont {I.}~\bibnamefont
  {Bloch}}, \bibinfo {author} {\bibfnamefont {J.}~\bibnamefont {Dalibard}},\
  and\ \bibinfo {author} {\bibfnamefont {S.}~\bibnamefont {Nascimb{\`e}ne}},\
  }\bibfield  {title} {\bibinfo {title} {Quantum simulations with ultracold
  quantum gases},\ }\href {https://doi.org/10.1038/nphys2259} {\bibfield
  {journal} {\bibinfo  {journal} {Nature Phys.}\ }\textbf {\bibinfo {volume}
  {8}},\ \bibinfo {pages} {267} (\bibinfo {year} {2012})}\BibitemShut {NoStop}%
\bibitem [{\citenamefont {De~Pasquale}\ \emph {et~al.}(2016)\citenamefont
  {De~Pasquale}, \citenamefont {Rossini}, \citenamefont {Fazio},\ and\
  \citenamefont {Giovannetti}}]{de2016local}%
  \BibitemOpen
  \bibfield  {author} {\bibinfo {author} {\bibfnamefont {A.}~\bibnamefont
  {De~Pasquale}}, \bibinfo {author} {\bibfnamefont {D.}~\bibnamefont
  {Rossini}}, \bibinfo {author} {\bibfnamefont {R.}~\bibnamefont {Fazio}},\
  and\ \bibinfo {author} {\bibfnamefont {V.}~\bibnamefont {Giovannetti}},\
  }\bibfield  {title} {\bibinfo {title} {Local quantum thermal
  susceptibility},\ }\href {https://doi.org/10.1038/ncomms12782} {\bibfield
  {journal} {\bibinfo  {journal} {Nat. Commun.}\ }\textbf {\bibinfo {volume}
  {7}},\ \bibinfo {pages} {12782} (\bibinfo {year} {2016})}\BibitemShut
  {NoStop}%
\bibitem [{\citenamefont {Mehboudi}\ \emph
  {et~al.}(2019{\natexlab{b}})\citenamefont {Mehboudi}, \citenamefont {Lampo},
  \citenamefont {Charalambous}, \citenamefont {Correa}, \citenamefont
  {Garc\'{\i}a-March},\ and\ \citenamefont
  {Lewenstein}}]{PhysRevLett.122.030403}%
  \BibitemOpen
  \bibfield  {author} {\bibinfo {author} {\bibfnamefont {M.}~\bibnamefont
  {Mehboudi}}, \bibinfo {author} {\bibfnamefont {A.}~\bibnamefont {Lampo}},
  \bibinfo {author} {\bibfnamefont {C.}~\bibnamefont {Charalambous}}, \bibinfo
  {author} {\bibfnamefont {L.~A.}\ \bibnamefont {Correa}}, \bibinfo {author}
  {\bibfnamefont {M.~A.}\ \bibnamefont {Garc\'{\i}a-March}},\ and\ \bibinfo
  {author} {\bibfnamefont {M.}~\bibnamefont {Lewenstein}},\ }\bibfield  {title}
  {\bibinfo {title} {Using polarons for sub-nk quantum nondemolition
  thermometry in a {B}ose-{E}instein condensate},\ }\href
  {https://doi.org/10.1103/PhysRevLett.122.030403} {\bibfield  {journal}
  {\bibinfo  {journal} {Phys. Rev. Lett.}\ }\textbf {\bibinfo {volume} {122}},\
  \bibinfo {pages} {030403} (\bibinfo {year} {2019}{\natexlab{b}})}\BibitemShut
  {NoStop}%
\bibitem [{\citenamefont {Hovhannisyan}\ and\ \citenamefont
  {Correa}(2018)}]{PhysRevB.98.045101}%
  \BibitemOpen
  \bibfield  {author} {\bibinfo {author} {\bibfnamefont {K.~V.}\ \bibnamefont
  {Hovhannisyan}}\ and\ \bibinfo {author} {\bibfnamefont {L.~A.}\ \bibnamefont
  {Correa}},\ }\bibfield  {title} {\bibinfo {title} {Measuring the temperature
  of cold many-body quantum systems},\ }\href
  {https://doi.org/10.1103/PhysRevB.98.045101} {\bibfield  {journal} {\bibinfo
  {journal} {Phys. Rev. B}\ }\textbf {\bibinfo {volume} {98}},\ \bibinfo
  {pages} {045101} (\bibinfo {year} {2018})}\BibitemShut {NoStop}%
\bibitem [{\citenamefont {Scigliuzzo}\ \emph {et~al.}(2020)\citenamefont
  {Scigliuzzo}, \citenamefont {Bengtsson}, \citenamefont {Besse}, \citenamefont
  {Wallraff}, \citenamefont {Delsing},\ and\ \citenamefont
  {Gasparinetti}}]{PhysRevX.10.041054}%
  \BibitemOpen
  \bibfield  {author} {\bibinfo {author} {\bibfnamefont {M.}~\bibnamefont
  {Scigliuzzo}}, \bibinfo {author} {\bibfnamefont {A.}~\bibnamefont
  {Bengtsson}}, \bibinfo {author} {\bibfnamefont {J.-C.}\ \bibnamefont
  {Besse}}, \bibinfo {author} {\bibfnamefont {A.}~\bibnamefont {Wallraff}},
  \bibinfo {author} {\bibfnamefont {P.}~\bibnamefont {Delsing}},\ and\ \bibinfo
  {author} {\bibfnamefont {S.}~\bibnamefont {Gasparinetti}},\ }\bibfield
  {title} {\bibinfo {title} {Primary thermometry of propagating microwaves in
  the quantum regime},\ }\href {https://doi.org/10.1103/PhysRevX.10.041054}
  {\bibfield  {journal} {\bibinfo  {journal} {Phys. Rev. X}\ }\textbf {\bibinfo
  {volume} {10}},\ \bibinfo {pages} {041054} (\bibinfo {year}
  {2020})}\BibitemShut {NoStop}%
\bibitem [{\citenamefont {Olf}\ \emph {et~al.}(2015)\citenamefont {Olf},
  \citenamefont {Fang}, \citenamefont {Marti}, \citenamefont {MacRae},\ and\
  \citenamefont {Stamper-Kurn}}]{olf2015thermometry}%
  \BibitemOpen
  \bibfield  {author} {\bibinfo {author} {\bibfnamefont {R.}~\bibnamefont
  {Olf}}, \bibinfo {author} {\bibfnamefont {F.}~\bibnamefont {Fang}}, \bibinfo
  {author} {\bibfnamefont {G.~E.}\ \bibnamefont {Marti}}, \bibinfo {author}
  {\bibfnamefont {A.}~\bibnamefont {MacRae}},\ and\ \bibinfo {author}
  {\bibfnamefont {D.~M.}\ \bibnamefont {Stamper-Kurn}},\ }\bibfield  {title}
  {\bibinfo {title} {Thermometry and cooling of a {B}ose gas to 0.02 times the
  condensation temperature},\ }\href {https://doi.org/10.1038/nphys3408}
  {\bibfield  {journal} {\bibinfo  {journal} {Nature Phys.}\ }\textbf {\bibinfo
  {volume} {11}},\ \bibinfo {pages} {720} (\bibinfo {year} {2015})}\BibitemShut
  {NoStop}%
\bibitem [{\citenamefont {Gati}\ \emph {et~al.}(2006)\citenamefont {Gati},
  \citenamefont {Hemmerling}, \citenamefont {F\"olling}, \citenamefont
  {Albiez},\ and\ \citenamefont {Oberthaler}}]{PhysRevLett.96.130404}%
  \BibitemOpen
  \bibfield  {author} {\bibinfo {author} {\bibfnamefont {R.}~\bibnamefont
  {Gati}}, \bibinfo {author} {\bibfnamefont {B.}~\bibnamefont {Hemmerling}},
  \bibinfo {author} {\bibfnamefont {J.}~\bibnamefont {F\"olling}}, \bibinfo
  {author} {\bibfnamefont {M.}~\bibnamefont {Albiez}},\ and\ \bibinfo {author}
  {\bibfnamefont {M.~K.}\ \bibnamefont {Oberthaler}},\ }\bibfield  {title}
  {\bibinfo {title} {Noise thermometry with two weakly coupled
  {B}ose-{E}instein condensates},\ }\href
  {https://doi.org/10.1103/PhysRevLett.96.130404} {\bibfield  {journal}
  {\bibinfo  {journal} {Phys. Rev. Lett.}\ }\textbf {\bibinfo {volume} {96}},\
  \bibinfo {pages} {130404} (\bibinfo {year} {2006})}\BibitemShut {NoStop}%
\bibitem [{\citenamefont {Mirkhalaf}\ \emph {et~al.}(2021)\citenamefont
  {Mirkhalaf}, \citenamefont {Benedicto~Orenes}, \citenamefont {Mitchell},\
  and\ \citenamefont {Witkowska}}]{PhysRevA.103.023317}%
  \BibitemOpen
  \bibfield  {author} {\bibinfo {author} {\bibfnamefont {S.~S.}\ \bibnamefont
  {Mirkhalaf}}, \bibinfo {author} {\bibfnamefont {D.}~\bibnamefont
  {Benedicto~Orenes}}, \bibinfo {author} {\bibfnamefont {M.~W.}\ \bibnamefont
  {Mitchell}},\ and\ \bibinfo {author} {\bibfnamefont {E.}~\bibnamefont
  {Witkowska}},\ }\bibfield  {title} {\bibinfo {title} {Criticality-enhanced
  quantum sensing in ferromagnetic {B}ose-{E}instein condensates: Role of
  readout measurement and detection noise},\ }\href
  {https://doi.org/10.1103/PhysRevA.103.023317} {\bibfield  {journal} {\bibinfo
   {journal} {Phys. Rev. A}\ }\textbf {\bibinfo {volume} {103}},\ \bibinfo
  {pages} {023317} (\bibinfo {year} {2021})}\BibitemShut {NoStop}%
\bibitem [{\citenamefont {Hofer}\ \emph {et~al.}(2017)\citenamefont {Hofer},
  \citenamefont {Brask}, \citenamefont {Perarnau-Llobet},\ and\ \citenamefont
  {Brunner}}]{PhysRevLett.119.090603}%
  \BibitemOpen
  \bibfield  {author} {\bibinfo {author} {\bibfnamefont {P.~P.}\ \bibnamefont
  {Hofer}}, \bibinfo {author} {\bibfnamefont {J.~B.}\ \bibnamefont {Brask}},
  \bibinfo {author} {\bibfnamefont {M.}~\bibnamefont {Perarnau-Llobet}},\ and\
  \bibinfo {author} {\bibfnamefont {N.}~\bibnamefont {Brunner}},\ }\bibfield
  {title} {\bibinfo {title} {Quantum thermal machine as a thermometer},\ }\href
  {https://doi.org/10.1103/PhysRevLett.119.090603} {\bibfield  {journal}
  {\bibinfo  {journal} {Phys. Rev. Lett.}\ }\textbf {\bibinfo {volume} {119}},\
  \bibinfo {pages} {090603} (\bibinfo {year} {2017})}\BibitemShut {NoStop}%
\bibitem [{\citenamefont {Mitchison}\ \emph {et~al.}(2020)\citenamefont
  {Mitchison}, \citenamefont {Fogarty}, \citenamefont {Guarnieri},
  \citenamefont {Campbell}, \citenamefont {Busch},\ and\ \citenamefont
  {Goold}}]{PhysRevLett.125.080402}%
  \BibitemOpen
  \bibfield  {author} {\bibinfo {author} {\bibfnamefont {M.~T.}\ \bibnamefont
  {Mitchison}}, \bibinfo {author} {\bibfnamefont {T.}~\bibnamefont {Fogarty}},
  \bibinfo {author} {\bibfnamefont {G.}~\bibnamefont {Guarnieri}}, \bibinfo
  {author} {\bibfnamefont {S.}~\bibnamefont {Campbell}}, \bibinfo {author}
  {\bibfnamefont {T.}~\bibnamefont {Busch}},\ and\ \bibinfo {author}
  {\bibfnamefont {J.}~\bibnamefont {Goold}},\ }\bibfield  {title} {\bibinfo
  {title} {In situ thermometry of a cold {F}ermi gas via dephasing
  impurities},\ }\href {https://doi.org/10.1103/PhysRevLett.125.080402}
  {\bibfield  {journal} {\bibinfo  {journal} {Phys. Rev. Lett.}\ }\textbf
  {\bibinfo {volume} {125}},\ \bibinfo {pages} {080402} (\bibinfo {year}
  {2020})}\BibitemShut {NoStop}%
\bibitem [{\citenamefont {Latune}\ \emph {et~al.}(2020)\citenamefont {Latune},
  \citenamefont {Sinayskiy},\ and\ \citenamefont
  {Petruccione}}]{latune2020collective}%
  \BibitemOpen
  \bibfield  {author} {\bibinfo {author} {\bibfnamefont {C.~L.}\ \bibnamefont
  {Latune}}, \bibinfo {author} {\bibfnamefont {I.}~\bibnamefont {Sinayskiy}},\
  and\ \bibinfo {author} {\bibfnamefont {F.}~\bibnamefont {Petruccione}},\
  }\bibfield  {title} {\bibinfo {title} {Collective heat capacity for quantum
  thermometry and quantum engine enhancements},\ }\href
  {https://doi.org/10.1088/1367-2630/aba463} {\bibfield  {journal} {\bibinfo
  {journal} {New J. Phys.}\ }\textbf {\bibinfo {volume} {22}},\ \bibinfo
  {pages} {083049} (\bibinfo {year} {2020})}\BibitemShut {NoStop}%
\bibitem [{\citenamefont {Rubio}\ \emph {et~al.}(2021)\citenamefont {Rubio},
  \citenamefont {Anders},\ and\ \citenamefont {Correa}}]{global_QT}%
  \BibitemOpen
  \bibfield  {author} {\bibinfo {author} {\bibfnamefont {J.}~\bibnamefont
  {Rubio}}, \bibinfo {author} {\bibfnamefont {J.}~\bibnamefont {Anders}},\ and\
  \bibinfo {author} {\bibfnamefont {L.~A.}\ \bibnamefont {Correa}},\ }\bibfield
   {title} {\bibinfo {title} {Global quantum thermometry},\ }\href
  {https://doi.org/10.1103/PhysRevLett.127.190402} {\bibfield  {journal}
  {\bibinfo  {journal} {Phys. Rev. Lett.}\ }\textbf {\bibinfo {volume} {127}},\
  \bibinfo {pages} {190402} (\bibinfo {year} {2021})}\BibitemShut {NoStop}%
\bibitem [{\citenamefont {Potts}\ \emph {et~al.}(2019)\citenamefont {Potts},
  \citenamefont {Brask},\ and\ \citenamefont {Brunner}}]{potts2019fundamental}%
  \BibitemOpen
  \bibfield  {author} {\bibinfo {author} {\bibfnamefont {P.~P.}\ \bibnamefont
  {Potts}}, \bibinfo {author} {\bibfnamefont {J.~B.}\ \bibnamefont {Brask}},\
  and\ \bibinfo {author} {\bibfnamefont {N.}~\bibnamefont {Brunner}},\
  }\bibfield  {title} {\bibinfo {title} {Fundamental limits on low-temperature
  quantum thermometry with finite resolution},\ }\href
  {https://doi.org/10.22331/q-2019-07-09-161} {\bibfield  {journal} {\bibinfo
  {journal} {{Quantum}}\ }\textbf {\bibinfo {volume} {3}},\ \bibinfo {pages}
  {161} (\bibinfo {year} {2019})}\BibitemShut {NoStop}%
\bibitem [{\citenamefont {Mehboudi}\ \emph {et~al.}(2015)\citenamefont
  {Mehboudi}, \citenamefont {Moreno-Cardoner}, \citenamefont {Chiara},\ and\
  \citenamefont {Sanpera}}]{mehboudi2015thermometry}%
  \BibitemOpen
  \bibfield  {author} {\bibinfo {author} {\bibfnamefont {M.}~\bibnamefont
  {Mehboudi}}, \bibinfo {author} {\bibfnamefont {M.}~\bibnamefont
  {Moreno-Cardoner}}, \bibinfo {author} {\bibfnamefont {G.~D.}\ \bibnamefont
  {Chiara}},\ and\ \bibinfo {author} {\bibfnamefont {A.}~\bibnamefont
  {Sanpera}},\ }\bibfield  {title} {\bibinfo {title} {Thermometry precision in
  strongly correlated ultracold lattice gases},\ }\href
  {https://doi.org/10.1088/1367-2630/17/5/055020} {\bibfield  {journal}
  {\bibinfo  {journal} {New J. Phys.}\ }\textbf {\bibinfo {volume} {17}},\
  \bibinfo {pages} {055020} (\bibinfo {year} {2015})}\BibitemShut {NoStop}%
\bibitem [{\citenamefont {Stace}(2010)}]{PhysRevA.82.011611}%
  \BibitemOpen
  \bibfield  {author} {\bibinfo {author} {\bibfnamefont {T.~M.}\ \bibnamefont
  {Stace}},\ }\bibfield  {title} {\bibinfo {title} {Quantum limits of
  thermometry},\ }\href {https://doi.org/10.1103/PhysRevA.82.011611} {\bibfield
   {journal} {\bibinfo  {journal} {Phys. Rev. A}\ }\textbf {\bibinfo {volume}
  {82}},\ \bibinfo {pages} {011611} (\bibinfo {year} {2010})}\BibitemShut
  {NoStop}%
\bibitem [{\citenamefont {Zhang}\ and\ \citenamefont {Tong}(2022)}]{Zhang2022}%
  \BibitemOpen
  \bibfield  {author} {\bibinfo {author} {\bibfnamefont {D.-J.}\ \bibnamefont
  {Zhang}}\ and\ \bibinfo {author} {\bibfnamefont {D.~M.}\ \bibnamefont
  {Tong}},\ }\bibfield  {title} {\bibinfo {title} {Approaching
  {H}eisenberg-scalable thermometry with built-in robustness against noise},\
  }\href {https://doi.org/10.1038/s41534-022-00588-2} {\bibfield  {journal}
  {\bibinfo  {journal} {npj Quantum Inf.}\ }\textbf {\bibinfo {volume} {8}},\
  \bibinfo {pages} {81} (\bibinfo {year} {2022})}\BibitemShut {NoStop}%
\bibitem [{\citenamefont {Rom{\'a}n-Ancheyta}\ \emph
  {et~al.}(2019)\citenamefont {Rom{\'a}n-Ancheyta}, \citenamefont
  {{\c{C}}akmak},\ and\ \citenamefont
  {M{\"u}stecapl{\i}o{\u{g}}lu}}]{Roman2019}%
  \BibitemOpen
  \bibfield  {author} {\bibinfo {author} {\bibfnamefont {R.}~\bibnamefont
  {Rom{\'a}n-Ancheyta}}, \bibinfo {author} {\bibfnamefont {B.}~\bibnamefont
  {{\c{C}}akmak}},\ and\ \bibinfo {author} {\bibfnamefont {{\"O}.~E.}\
  \bibnamefont {M{\"u}stecapl{\i}o{\u{g}}lu}},\ }\bibfield  {title} {\bibinfo
  {title} {Spectral signatures of non-thermal baths in quantum
  thermalization},\ }\href {https://doi.org/10.1088/2058-9565/ab5e4f}
  {\bibfield  {journal} {\bibinfo  {journal} {Quantum Sci.Technol.}\ }\textbf
  {\bibinfo {volume} {5}},\ \bibinfo {pages} {015003} (\bibinfo {year}
  {2019})}\BibitemShut {NoStop}%
\bibitem [{\citenamefont {Brunelli}\ \emph {et~al.}(2012)\citenamefont
  {Brunelli}, \citenamefont {Olivares}, \citenamefont {Paternostro},\ and\
  \citenamefont {Paris}}]{PhysRevA.86.012125}%
  \BibitemOpen
  \bibfield  {author} {\bibinfo {author} {\bibfnamefont {M.}~\bibnamefont
  {Brunelli}}, \bibinfo {author} {\bibfnamefont {S.}~\bibnamefont {Olivares}},
  \bibinfo {author} {\bibfnamefont {M.}~\bibnamefont {Paternostro}},\ and\
  \bibinfo {author} {\bibfnamefont {M.~G.~A.}\ \bibnamefont {Paris}},\
  }\bibfield  {title} {\bibinfo {title} {Qubit-assisted thermometry of a
  quantum harmonic oscillator},\ }\href
  {https://doi.org/10.1103/PhysRevA.86.012125} {\bibfield  {journal} {\bibinfo
  {journal} {Phys. Rev. A}\ }\textbf {\bibinfo {volume} {86}},\ \bibinfo
  {pages} {012125} (\bibinfo {year} {2012})}\BibitemShut {NoStop}%
\bibitem [{\citenamefont {Brunelli}\ \emph {et~al.}(2011)\citenamefont
  {Brunelli}, \citenamefont {Olivares},\ and\ \citenamefont
  {Paris}}]{PhysRevA.84.032105}%
  \BibitemOpen
  \bibfield  {author} {\bibinfo {author} {\bibfnamefont {M.}~\bibnamefont
  {Brunelli}}, \bibinfo {author} {\bibfnamefont {S.}~\bibnamefont {Olivares}},\
  and\ \bibinfo {author} {\bibfnamefont {M.~G.~A.}\ \bibnamefont {Paris}},\
  }\bibfield  {title} {\bibinfo {title} {Qubit thermometry for micromechanical
  resonators},\ }\href {https://doi.org/10.1103/PhysRevA.84.032105} {\bibfield
  {journal} {\bibinfo  {journal} {Phys. Rev. A}\ }\textbf {\bibinfo {volume}
  {84}},\ \bibinfo {pages} {032105} (\bibinfo {year} {2011})}\BibitemShut
  {NoStop}%
\bibitem [{\citenamefont {Jevtic}\ \emph {et~al.}(2015)\citenamefont {Jevtic},
  \citenamefont {Newman}, \citenamefont {Rudolph},\ and\ \citenamefont
  {Stace}}]{PhysRevA.91.012331}%
  \BibitemOpen
  \bibfield  {author} {\bibinfo {author} {\bibfnamefont {S.}~\bibnamefont
  {Jevtic}}, \bibinfo {author} {\bibfnamefont {D.}~\bibnamefont {Newman}},
  \bibinfo {author} {\bibfnamefont {T.}~\bibnamefont {Rudolph}},\ and\ \bibinfo
  {author} {\bibfnamefont {T.~M.}\ \bibnamefont {Stace}},\ }\bibfield  {title}
  {\bibinfo {title} {Single-qubit thermometry},\ }\href
  {https://doi.org/10.1103/PhysRevA.91.012331} {\bibfield  {journal} {\bibinfo
  {journal} {Phys. Rev. A}\ }\textbf {\bibinfo {volume} {91}},\ \bibinfo
  {pages} {012331} (\bibinfo {year} {2015})}\BibitemShut {NoStop}%
\bibitem [{\citenamefont {Mancino}\ \emph {et~al.}(2017)\citenamefont
  {Mancino}, \citenamefont {Sbroscia}, \citenamefont {Gianani}, \citenamefont
  {Roccia},\ and\ \citenamefont {Barbieri}}]{PhysRevLett.118.130502}%
  \BibitemOpen
  \bibfield  {author} {\bibinfo {author} {\bibfnamefont {L.}~\bibnamefont
  {Mancino}}, \bibinfo {author} {\bibfnamefont {M.}~\bibnamefont {Sbroscia}},
  \bibinfo {author} {\bibfnamefont {I.}~\bibnamefont {Gianani}}, \bibinfo
  {author} {\bibfnamefont {E.}~\bibnamefont {Roccia}},\ and\ \bibinfo {author}
  {\bibfnamefont {M.}~\bibnamefont {Barbieri}},\ }\bibfield  {title} {\bibinfo
  {title} {Quantum simulation of single-qubit thermometry using linear
  optics},\ }\href {https://doi.org/10.1103/PhysRevLett.118.130502} {\bibfield
  {journal} {\bibinfo  {journal} {Phys. Rev. Lett.}\ }\textbf {\bibinfo
  {volume} {118}},\ \bibinfo {pages} {130502} (\bibinfo {year}
  {2017})}\BibitemShut {NoStop}%
\bibitem [{\citenamefont {Kiilerich}\ \emph {et~al.}(2018)\citenamefont
  {Kiilerich}, \citenamefont {De~Pasquale},\ and\ \citenamefont
  {Giovannetti}}]{PhysRevA.98.042124}%
  \BibitemOpen
  \bibfield  {author} {\bibinfo {author} {\bibfnamefont {A.~H.}\ \bibnamefont
  {Kiilerich}}, \bibinfo {author} {\bibfnamefont {A.}~\bibnamefont
  {De~Pasquale}},\ and\ \bibinfo {author} {\bibfnamefont {V.}~\bibnamefont
  {Giovannetti}},\ }\bibfield  {title} {\bibinfo {title} {Dynamical approach to
  ancilla-assisted quantum thermometry},\ }\href
  {https://doi.org/10.1103/PhysRevA.98.042124} {\bibfield  {journal} {\bibinfo
  {journal} {Phys. Rev. A}\ }\textbf {\bibinfo {volume} {98}},\ \bibinfo
  {pages} {042124} (\bibinfo {year} {2018})}\BibitemShut {NoStop}%
\bibitem [{\citenamefont {Feyles}\ \emph {et~al.}(2019)\citenamefont {Feyles},
  \citenamefont {Mancino}, \citenamefont {Sbroscia}, \citenamefont {Gianani},\
  and\ \citenamefont {Barbieri}}]{PhysRevA.99.062114}%
  \BibitemOpen
  \bibfield  {author} {\bibinfo {author} {\bibfnamefont {M.~M.}\ \bibnamefont
  {Feyles}}, \bibinfo {author} {\bibfnamefont {L.}~\bibnamefont {Mancino}},
  \bibinfo {author} {\bibfnamefont {M.}~\bibnamefont {Sbroscia}}, \bibinfo
  {author} {\bibfnamefont {I.}~\bibnamefont {Gianani}},\ and\ \bibinfo {author}
  {\bibfnamefont {M.}~\bibnamefont {Barbieri}},\ }\bibfield  {title} {\bibinfo
  {title} {Dynamical role of quantum signatures in quantum thermometry},\
  }\href {https://doi.org/10.1103/PhysRevA.99.062114} {\bibfield  {journal}
  {\bibinfo  {journal} {Phys. Rev. A}\ }\textbf {\bibinfo {volume} {99}},\
  \bibinfo {pages} {062114} (\bibinfo {year} {2019})}\BibitemShut {NoStop}%
\bibitem [{\citenamefont {Gianani}\ \emph {et~al.}(2020)\citenamefont
  {Gianani}, \citenamefont {Farina}, \citenamefont {Barbieri}, \citenamefont
  {Cimini}, \citenamefont {Cavina},\ and\ \citenamefont
  {Giovannetti}}]{PhysRevResearch.2.033497}%
  \BibitemOpen
  \bibfield  {author} {\bibinfo {author} {\bibfnamefont {I.}~\bibnamefont
  {Gianani}}, \bibinfo {author} {\bibfnamefont {D.}~\bibnamefont {Farina}},
  \bibinfo {author} {\bibfnamefont {M.}~\bibnamefont {Barbieri}}, \bibinfo
  {author} {\bibfnamefont {V.}~\bibnamefont {Cimini}}, \bibinfo {author}
  {\bibfnamefont {V.}~\bibnamefont {Cavina}},\ and\ \bibinfo {author}
  {\bibfnamefont {V.}~\bibnamefont {Giovannetti}},\ }\bibfield  {title}
  {\bibinfo {title} {Discrimination of thermal baths by single-qubit probes},\
  }\href {https://doi.org/10.1103/PhysRevResearch.2.033497} {\bibfield
  {journal} {\bibinfo  {journal} {Phys. Rev. Research}\ }\textbf {\bibinfo
  {volume} {2}},\ \bibinfo {pages} {033497} (\bibinfo {year}
  {2020})}\BibitemShut {NoStop}%
\bibitem [{\citenamefont {J\o{}rgensen}\ \emph {et~al.}(2020)\citenamefont
  {J\o{}rgensen}, \citenamefont {Potts}, \citenamefont {Paris},\ and\
  \citenamefont {Brask}}]{PhysRevResearch.2.033394}%
  \BibitemOpen
  \bibfield  {author} {\bibinfo {author} {\bibfnamefont {M.~R.}\ \bibnamefont
  {J\o{}rgensen}}, \bibinfo {author} {\bibfnamefont {P.~P.}\ \bibnamefont
  {Potts}}, \bibinfo {author} {\bibfnamefont {M.~G.~A.}\ \bibnamefont
  {Paris}},\ and\ \bibinfo {author} {\bibfnamefont {J.~B.}\ \bibnamefont
  {Brask}},\ }\bibfield  {title} {\bibinfo {title} {Tight bound on
  finite-resolution quantum thermometry at low temperatures},\ }\href
  {https://doi.org/10.1103/PhysRevResearch.2.033394} {\bibfield  {journal}
  {\bibinfo  {journal} {Phys. Rev. Research}\ }\textbf {\bibinfo {volume}
  {2}},\ \bibinfo {pages} {033394} (\bibinfo {year} {2020})}\BibitemShut
  {NoStop}%
\bibitem [{\citenamefont {Bouton}\ \emph {et~al.}(2020)\citenamefont {Bouton},
  \citenamefont {Nettersheim}, \citenamefont {Adam}, \citenamefont {Schmidt},
  \citenamefont {Mayer}, \citenamefont {Lausch}, \citenamefont {Tiemann},\ and\
  \citenamefont {Widera}}]{PhysRevX.10.011018}%
  \BibitemOpen
  \bibfield  {author} {\bibinfo {author} {\bibfnamefont {Q.}~\bibnamefont
  {Bouton}}, \bibinfo {author} {\bibfnamefont {J.}~\bibnamefont {Nettersheim}},
  \bibinfo {author} {\bibfnamefont {D.}~\bibnamefont {Adam}}, \bibinfo {author}
  {\bibfnamefont {F.}~\bibnamefont {Schmidt}}, \bibinfo {author} {\bibfnamefont
  {D.}~\bibnamefont {Mayer}}, \bibinfo {author} {\bibfnamefont
  {T.}~\bibnamefont {Lausch}}, \bibinfo {author} {\bibfnamefont
  {E.}~\bibnamefont {Tiemann}},\ and\ \bibinfo {author} {\bibfnamefont
  {A.}~\bibnamefont {Widera}},\ }\bibfield  {title} {\bibinfo {title}
  {Single-atom quantum probes for ultracold gases boosted by nonequilibrium
  spin dynamics},\ }\href {https://doi.org/10.1103/PhysRevX.10.011018}
  {\bibfield  {journal} {\bibinfo  {journal} {Phys. Rev. X}\ }\textbf {\bibinfo
  {volume} {10}},\ \bibinfo {pages} {011018} (\bibinfo {year}
  {2020})}\BibitemShut {NoStop}%
\bibitem [{\citenamefont {Campbell}\ \emph {et~al.}(2018)\citenamefont
  {Campbell}, \citenamefont {Genoni},\ and\ \citenamefont
  {Deffner}}]{campbell2018precision}%
  \BibitemOpen
  \bibfield  {author} {\bibinfo {author} {\bibfnamefont {S.}~\bibnamefont
  {Campbell}}, \bibinfo {author} {\bibfnamefont {M.~G.}\ \bibnamefont
  {Genoni}},\ and\ \bibinfo {author} {\bibfnamefont {S.}~\bibnamefont
  {Deffner}},\ }\bibfield  {title} {\bibinfo {title} {Precision thermometry and
  the quantum speed limit},\ }\href {https://doi.org/10.1088/2058-9565/aaa641}
  {\bibfield  {journal} {\bibinfo  {journal} {Quantum Sci. Technol.}\ }\textbf
  {\bibinfo {volume} {3}},\ \bibinfo {pages} {025002} (\bibinfo {year}
  {2018})}\BibitemShut {NoStop}%
\bibitem [{\citenamefont {Correa}\ \emph {et~al.}(2015)\citenamefont {Correa},
  \citenamefont {Mehboudi}, \citenamefont {Adesso},\ and\ \citenamefont
  {Sanpera}}]{PhysRevLett.114.220405}%
  \BibitemOpen
  \bibfield  {author} {\bibinfo {author} {\bibfnamefont {L.~A.}\ \bibnamefont
  {Correa}}, \bibinfo {author} {\bibfnamefont {M.}~\bibnamefont {Mehboudi}},
  \bibinfo {author} {\bibfnamefont {G.}~\bibnamefont {Adesso}},\ and\ \bibinfo
  {author} {\bibfnamefont {A.}~\bibnamefont {Sanpera}},\ }\bibfield  {title}
  {\bibinfo {title} {Individual quantum probes for optimal thermometry},\
  }\href {https://doi.org/10.1103/PhysRevLett.114.220405} {\bibfield  {journal}
  {\bibinfo  {journal} {Phys. Rev. Lett.}\ }\textbf {\bibinfo {volume} {114}},\
  \bibinfo {pages} {220405} (\bibinfo {year} {2015})}\BibitemShut {NoStop}%
\bibitem [{\citenamefont {Mok}\ \emph {et~al.}(2021)\citenamefont {Mok},
  \citenamefont {Bharti}, \citenamefont {Kwek},\ and\ \citenamefont
  {Bayat}}]{Mok2021}%
  \BibitemOpen
  \bibfield  {author} {\bibinfo {author} {\bibfnamefont {W.-K.}\ \bibnamefont
  {Mok}}, \bibinfo {author} {\bibfnamefont {K.}~\bibnamefont {Bharti}},
  \bibinfo {author} {\bibfnamefont {L.-C.}\ \bibnamefont {Kwek}},\ and\
  \bibinfo {author} {\bibfnamefont {A.}~\bibnamefont {Bayat}},\ }\bibfield
  {title} {\bibinfo {title} {Optimal probes for global quantum thermometry},\
  }\href {https://doi.org/10.1038/s42005-021-00572-w} {\bibfield  {journal}
  {\bibinfo  {journal} {Commun. Phys.}\ }\textbf {\bibinfo {volume} {4}},\
  \bibinfo {pages} {62} (\bibinfo {year} {2021})}\BibitemShut {NoStop}%
\bibitem [{\citenamefont {Boeyens}\ \emph {et~al.}(2021)\citenamefont
  {Boeyens}, \citenamefont {Seah},\ and\ \citenamefont
  {Nimmrichter}}]{PhysRevA.104.052214}%
  \BibitemOpen
  \bibfield  {author} {\bibinfo {author} {\bibfnamefont {J.}~\bibnamefont
  {Boeyens}}, \bibinfo {author} {\bibfnamefont {S.}~\bibnamefont {Seah}},\ and\
  \bibinfo {author} {\bibfnamefont {S.}~\bibnamefont {Nimmrichter}},\
  }\bibfield  {title} {\bibinfo {title} {Uninformed {B}ayesian quantum
  thermometry},\ }\href {https://doi.org/10.1103/PhysRevA.104.052214}
  {\bibfield  {journal} {\bibinfo  {journal} {Phys. Rev. A}\ }\textbf {\bibinfo
  {volume} {104}},\ \bibinfo {pages} {052214} (\bibinfo {year}
  {2021})}\BibitemShut {NoStop}%
\bibitem [{\citenamefont {J\o{}rgensen}\ \emph {et~al.}(2022)\citenamefont
  {J\o{}rgensen}, \citenamefont {Ko\l{}ody\ifmmode~\acute{n}\else
  \'{n}\fi{}ski}, \citenamefont {Mehboudi}, \citenamefont {Perarnau-Llobet},\
  and\ \citenamefont {Brask}}]{PhysRevA.105.042601}%
  \BibitemOpen
  \bibfield  {author} {\bibinfo {author} {\bibfnamefont {M.~R.}\ \bibnamefont
  {J\o{}rgensen}}, \bibinfo {author} {\bibfnamefont {J.}~\bibnamefont
  {Ko\l{}ody\ifmmode~\acute{n}\else \'{n}\fi{}ski}}, \bibinfo {author}
  {\bibfnamefont {M.}~\bibnamefont {Mehboudi}}, \bibinfo {author}
  {\bibfnamefont {M.}~\bibnamefont {Perarnau-Llobet}},\ and\ \bibinfo {author}
  {\bibfnamefont {J.~B.}\ \bibnamefont {Brask}},\ }\bibfield  {title} {\bibinfo
  {title} {Bayesian quantum thermometry based on thermodynamic length},\ }\href
  {https://doi.org/10.1103/PhysRevA.105.042601} {\bibfield  {journal} {\bibinfo
   {journal} {Phys. Rev. A}\ }\textbf {\bibinfo {volume} {105}},\ \bibinfo
  {pages} {042601} (\bibinfo {year} {2022})}\BibitemShut {NoStop}%
\bibitem [{\citenamefont {Alves}\ and\ \citenamefont
  {Landi}(2022)}]{PhysRevA.105.012212}%
  \BibitemOpen
  \bibfield  {author} {\bibinfo {author} {\bibfnamefont {G.~O.}\ \bibnamefont
  {Alves}}\ and\ \bibinfo {author} {\bibfnamefont {G.~T.}\ \bibnamefont
  {Landi}},\ }\bibfield  {title} {\bibinfo {title} {Bayesian estimation for
  collisional thermometry},\ }\href
  {https://doi.org/10.1103/PhysRevA.105.012212} {\bibfield  {journal} {\bibinfo
   {journal} {Phys. Rev. A}\ }\textbf {\bibinfo {volume} {105}},\ \bibinfo
  {pages} {012212} (\bibinfo {year} {2022})}\BibitemShut {NoStop}%
\bibitem [{\citenamefont {Mehboudi}\ \emph {et~al.}(2022)\citenamefont
  {Mehboudi}, \citenamefont {J\o{}rgensen}, \citenamefont {Seah}, \citenamefont
  {Brask}, \citenamefont {Ko\l{}ody\ifmmode~\acute{n}\else \'{n}\fi{}ski},\
  and\ \citenamefont {Perarnau-Llobet}}]{PhysRevLett.128.130502}%
  \BibitemOpen
  \bibfield  {author} {\bibinfo {author} {\bibfnamefont {M.}~\bibnamefont
  {Mehboudi}}, \bibinfo {author} {\bibfnamefont {M.~R.}\ \bibnamefont
  {J\o{}rgensen}}, \bibinfo {author} {\bibfnamefont {S.}~\bibnamefont {Seah}},
  \bibinfo {author} {\bibfnamefont {J.~B.}\ \bibnamefont {Brask}}, \bibinfo
  {author} {\bibfnamefont {J.}~\bibnamefont {Ko\l{}ody\ifmmode~\acute{n}\else
  \'{n}\fi{}ski}},\ and\ \bibinfo {author} {\bibfnamefont {M.}~\bibnamefont
  {Perarnau-Llobet}},\ }\bibfield  {title} {\bibinfo {title} {Fundamental
  limits in {B}ayesian thermometry and attainability via adaptive strategies},\
  }\href {https://doi.org/10.1103/PhysRevLett.128.130502} {\bibfield  {journal}
  {\bibinfo  {journal} {Phys. Rev. Lett.}\ }\textbf {\bibinfo {volume} {128}},\
  \bibinfo {pages} {130502} (\bibinfo {year} {2022})}\BibitemShut {NoStop}%
\bibitem [{\citenamefont {Glatthard}\ \emph {et~al.}(2022)\citenamefont
  {Glatthard}, \citenamefont {Rubio}, \citenamefont {Sawant}, \citenamefont
  {Hewitt}, \citenamefont {Barontini},\ and\ \citenamefont
  {Correa}}]{PRXQuantum.3.040330}%
  \BibitemOpen
  \bibfield  {author} {\bibinfo {author} {\bibfnamefont {J.}~\bibnamefont
  {Glatthard}}, \bibinfo {author} {\bibfnamefont {J.}~\bibnamefont {Rubio}},
  \bibinfo {author} {\bibfnamefont {R.}~\bibnamefont {Sawant}}, \bibinfo
  {author} {\bibfnamefont {T.}~\bibnamefont {Hewitt}}, \bibinfo {author}
  {\bibfnamefont {G.}~\bibnamefont {Barontini}},\ and\ \bibinfo {author}
  {\bibfnamefont {L.~A.}\ \bibnamefont {Correa}},\ }\bibfield  {title}
  {\bibinfo {title} {Optimal cold atom thermometry using adaptive {B}ayesian
  strategies},\ }\href {https://doi.org/10.1103/PRXQuantum.3.040330} {\bibfield
   {journal} {\bibinfo  {journal} {PRX Quantum}\ }\textbf {\bibinfo {volume}
  {3}},\ \bibinfo {pages} {040330} (\bibinfo {year} {2022})}\BibitemShut
  {NoStop}%
\bibitem [{\citenamefont {Correa}\ \emph {et~al.}(2017)\citenamefont {Correa},
  \citenamefont {Perarnau-Llobet}, \citenamefont {Hovhannisyan}, \citenamefont
  {Hern\'andez-Santana}, \citenamefont {Mehboudi},\ and\ \citenamefont
  {Sanpera}}]{PhysRevA.96.062103}%
  \BibitemOpen
  \bibfield  {author} {\bibinfo {author} {\bibfnamefont {L.~A.}\ \bibnamefont
  {Correa}}, \bibinfo {author} {\bibfnamefont {M.}~\bibnamefont
  {Perarnau-Llobet}}, \bibinfo {author} {\bibfnamefont {K.~V.}\ \bibnamefont
  {Hovhannisyan}}, \bibinfo {author} {\bibfnamefont {S.}~\bibnamefont
  {Hern\'andez-Santana}}, \bibinfo {author} {\bibfnamefont {M.}~\bibnamefont
  {Mehboudi}},\ and\ \bibinfo {author} {\bibfnamefont {A.}~\bibnamefont
  {Sanpera}},\ }\bibfield  {title} {\bibinfo {title} {Enhancement of
  low-temperature thermometry by strong coupling},\ }\href
  {https://doi.org/10.1103/PhysRevA.96.062103} {\bibfield  {journal} {\bibinfo
  {journal} {Phys. Rev. A}\ }\textbf {\bibinfo {volume} {96}},\ \bibinfo
  {pages} {062103} (\bibinfo {year} {2017})}\BibitemShut {NoStop}%
\bibitem [{\citenamefont {Glatthard}\ and\ \citenamefont
  {Correa}(2022)}]{glatthard2022bending}%
  \BibitemOpen
  \bibfield  {author} {\bibinfo {author} {\bibfnamefont {J.}~\bibnamefont
  {Glatthard}}\ and\ \bibinfo {author} {\bibfnamefont {L.~A.}\ \bibnamefont
  {Correa}},\ }\bibfield  {title} {\bibinfo {title} {Bending the rules of
  low-temperature thermometry with periodic driving},\ }\href
  {https://doi.org/10.22331/q-2022-05-03-705} {\bibfield  {journal} {\bibinfo
  {journal} {{Quantum}}\ }\textbf {\bibinfo {volume} {6}},\ \bibinfo {pages}
  {705} (\bibinfo {year} {2022})}\BibitemShut {NoStop}%
\bibitem [{\citenamefont {Mukherjee}\ \emph {et~al.}(2019)\citenamefont
  {Mukherjee}, \citenamefont {Zwick}, \citenamefont {Ghosh}, \citenamefont
  {Chen},\ and\ \citenamefont {Kurizki}}]{Mukherjee2019}%
  \BibitemOpen
  \bibfield  {author} {\bibinfo {author} {\bibfnamefont {V.}~\bibnamefont
  {Mukherjee}}, \bibinfo {author} {\bibfnamefont {A.}~\bibnamefont {Zwick}},
  \bibinfo {author} {\bibfnamefont {A.}~\bibnamefont {Ghosh}}, \bibinfo
  {author} {\bibfnamefont {X.}~\bibnamefont {Chen}},\ and\ \bibinfo {author}
  {\bibfnamefont {G.}~\bibnamefont {Kurizki}},\ }\bibfield  {title} {\bibinfo
  {title} {Enhanced precision bound of low-temperature quantum thermometry via
  dynamical control},\ }\href {https://doi.org/10.1038/s42005-019-0265-y}
  {\bibfield  {journal} {\bibinfo  {journal} {Commun. Phys.}\ }\textbf
  {\bibinfo {volume} {2}},\ \bibinfo {pages} {162} (\bibinfo {year}
  {2019})}\BibitemShut {NoStop}%
\bibitem [{\citenamefont {Kosloff}(2013)}]{e15062100}%
  \BibitemOpen
  \bibfield  {author} {\bibinfo {author} {\bibfnamefont {R.}~\bibnamefont
  {Kosloff}},\ }\bibfield  {title} {\bibinfo {title} {Quantum thermodynamics: A
  dynamical viewpoint},\ }\href {https://doi.org/10.3390/e15062100} {\bibfield
  {journal} {\bibinfo  {journal} {Entropy}\ }\textbf {\bibinfo {volume} {15}},\
  \bibinfo {pages} {2100} (\bibinfo {year} {2013})}\BibitemShut {NoStop}%
\bibitem [{\citenamefont {Binder}\ \emph {et~al.}(2018)\citenamefont {Binder},
  \citenamefont {Correa}, \citenamefont {Gogolin}, \citenamefont {Anders},\
  and\ \citenamefont {Adesso}}]{binder2018thermodynamics}%
  \BibitemOpen
  \bibfield  {author} {\bibinfo {author} {\bibfnamefont {F.}~\bibnamefont
  {Binder}}, \bibinfo {author} {\bibfnamefont {L.~A.}\ \bibnamefont {Correa}},
  \bibinfo {author} {\bibfnamefont {C.}~\bibnamefont {Gogolin}}, \bibinfo
  {author} {\bibfnamefont {J.}~\bibnamefont {Anders}},\ and\ \bibinfo {author}
  {\bibfnamefont {G.}~\bibnamefont {Adesso}},\ }\bibfield  {title} {\bibinfo
  {title} {Thermodynamics in the quantum regime},\ }\href@noop {} {\bibfield
  {journal} {\bibinfo  {journal} {Fundamental Theories of Physics}\ }\textbf
  {\bibinfo {volume} {195}},\ \bibinfo {pages} {1} (\bibinfo {year}
  {2018})}\BibitemShut {NoStop}%
\bibitem [{\citenamefont {Deffner}\ and\ \citenamefont
  {Campbell}(2019)}]{campbellbook}%
  \BibitemOpen
  \bibfield  {author} {\bibinfo {author} {\bibfnamefont {S.}~\bibnamefont
  {Deffner}}\ and\ \bibinfo {author} {\bibfnamefont {S.}~\bibnamefont
  {Campbell}},\ }\href {https://doi.org/10.1088/2053-2571/ab21c6} {\emph
  {\bibinfo {title} {Quantum Thermodynamics}}}\ (\bibinfo  {publisher} {(Morgan
  \& Claypool Publishers, San Rafael, CA},\ \bibinfo {year} {2019})\BibitemShut
  {NoStop}%
\bibitem [{\citenamefont {A.}\ and\ \citenamefont {M\"ustecapl\ifmmode \imath
  \else \i \fi{}o\ifmmode~\breve{g}\else \u{g}\fi{}lu}(2020)}]{AsliReview}%
  \BibitemOpen
  \bibfield  {author} {\bibinfo {author} {\bibfnamefont {T.}~\bibnamefont
  {A.}}\ and\ \bibinfo {author} {\bibfnamefont {O.~E.}\ \bibnamefont
  {M\"ustecapl\ifmmode \imath \else \i \fi{}o\ifmmode~\breve{g}\else
  \u{g}\fi{}lu}},\ }\bibfield  {title} {\bibinfo {title} {Quantum
  thermodynamics and quantum coherence engines},\ }\href
  {https://doi.org/10.3906/fiz-2009-12} {\bibfield  {journal} {\bibinfo
  {journal} {Turk. J. Phys.}\ }\textbf {\bibinfo {volume} {44}},\ \bibinfo
  {pages} {404} (\bibinfo {year} {2020})}\BibitemShut {NoStop}%
\bibitem [{\citenamefont {Kurizki}\ and\ \citenamefont
  {Kofman}(2022)}]{kurizki_kofman_2022}%
  \BibitemOpen
  \bibfield  {author} {\bibinfo {author} {\bibfnamefont {G.}~\bibnamefont
  {Kurizki}}\ and\ \bibinfo {author} {\bibfnamefont {A.~G.}\ \bibnamefont
  {Kofman}},\ }\href {https://doi.org/10.1017/9781316798454} {\emph {\bibinfo
  {title} {Thermodynamics and Control of Open Quantum Systems}}}\ (\bibinfo
  {publisher} {Cambridge University Press, Cambridge},\ \bibinfo {year}
  {2022})\BibitemShut {NoStop}%
\bibitem [{\citenamefont {Gebbia}\ \emph {et~al.}(2020)\citenamefont {Gebbia},
  \citenamefont {Benedetti}, \citenamefont {Benatti}, \citenamefont
  {Floreanini}, \citenamefont {Bina},\ and\ \citenamefont
  {Paris}}]{PhysRevA.101.032112}%
  \BibitemOpen
  \bibfield  {author} {\bibinfo {author} {\bibfnamefont {F.}~\bibnamefont
  {Gebbia}}, \bibinfo {author} {\bibfnamefont {C.}~\bibnamefont {Benedetti}},
  \bibinfo {author} {\bibfnamefont {F.}~\bibnamefont {Benatti}}, \bibinfo
  {author} {\bibfnamefont {R.}~\bibnamefont {Floreanini}}, \bibinfo {author}
  {\bibfnamefont {M.}~\bibnamefont {Bina}},\ and\ \bibinfo {author}
  {\bibfnamefont {M.~G.~A.}\ \bibnamefont {Paris}},\ }\bibfield  {title}
  {\bibinfo {title} {Two-qubit quantum probes for the temperature of an ohmic
  environment},\ }\href {https://doi.org/10.1103/PhysRevA.101.032112}
  {\bibfield  {journal} {\bibinfo  {journal} {Phys. Rev. A}\ }\textbf {\bibinfo
  {volume} {101}},\ \bibinfo {pages} {032112} (\bibinfo {year}
  {2020})}\BibitemShut {NoStop}%
\bibitem [{\citenamefont {Seah}\ \emph {et~al.}(2019)\citenamefont {Seah},
  \citenamefont {Nimmrichter}, \citenamefont {Grimmer}, \citenamefont {Santos},
  \citenamefont {Scarani},\ and\ \citenamefont
  {Landi}}]{PhysRevLett.123.180602}%
  \BibitemOpen
  \bibfield  {author} {\bibinfo {author} {\bibfnamefont {S.}~\bibnamefont
  {Seah}}, \bibinfo {author} {\bibfnamefont {S.}~\bibnamefont {Nimmrichter}},
  \bibinfo {author} {\bibfnamefont {D.}~\bibnamefont {Grimmer}}, \bibinfo
  {author} {\bibfnamefont {J.~P.}\ \bibnamefont {Santos}}, \bibinfo {author}
  {\bibfnamefont {V.}~\bibnamefont {Scarani}},\ and\ \bibinfo {author}
  {\bibfnamefont {G.~T.}\ \bibnamefont {Landi}},\ }\bibfield  {title} {\bibinfo
  {title} {Collisional quantum thermometry},\ }\href
  {https://doi.org/10.1103/PhysRevLett.123.180602} {\bibfield  {journal}
  {\bibinfo  {journal} {Phys. Rev. Lett.}\ }\textbf {\bibinfo {volume} {123}},\
  \bibinfo {pages} {180602} (\bibinfo {year} {2019})}\BibitemShut {NoStop}%
\bibitem [{\citenamefont {Campbell}\ \emph {et~al.}(2017)\citenamefont
  {Campbell}, \citenamefont {Mehboudi}, \citenamefont {Chiara},\ and\
  \citenamefont {Paternostro}}]{Campbell2017}%
  \BibitemOpen
  \bibfield  {author} {\bibinfo {author} {\bibfnamefont {S.}~\bibnamefont
  {Campbell}}, \bibinfo {author} {\bibfnamefont {M.}~\bibnamefont {Mehboudi}},
  \bibinfo {author} {\bibfnamefont {G.~D.}\ \bibnamefont {Chiara}},\ and\
  \bibinfo {author} {\bibfnamefont {M.}~\bibnamefont {Paternostro}},\
  }\bibfield  {title} {\bibinfo {title} {Global and local thermometry schemes
  in coupled quantum systems},\ }\href
  {https://doi.org/10.1088/1367-2630/aa7fac} {\bibfield  {journal} {\bibinfo
  {journal} {New J. Phys.}\ }\textbf {\bibinfo {volume} {19}},\ \bibinfo
  {pages} {103003} (\bibinfo {year} {2017})}\BibitemShut {NoStop}%
\bibitem [{\citenamefont {De~Pasquale}\ \emph {et~al.}(2017)\citenamefont
  {De~Pasquale}, \citenamefont {Yuasa},\ and\ \citenamefont
  {Giovannetti}}]{PhysRevA.96.012316}%
  \BibitemOpen
  \bibfield  {author} {\bibinfo {author} {\bibfnamefont {A.}~\bibnamefont
  {De~Pasquale}}, \bibinfo {author} {\bibfnamefont {K.}~\bibnamefont {Yuasa}},\
  and\ \bibinfo {author} {\bibfnamefont {V.}~\bibnamefont {Giovannetti}},\
  }\bibfield  {title} {\bibinfo {title} {Estimating temperature via sequential
  measurements},\ }\href {https://doi.org/10.1103/PhysRevA.96.012316}
  {\bibfield  {journal} {\bibinfo  {journal} {Phys. Rev. A}\ }\textbf {\bibinfo
  {volume} {96}},\ \bibinfo {pages} {012316} (\bibinfo {year}
  {2017})}\BibitemShut {NoStop}%
\bibitem [{\citenamefont {Cavina}\ \emph {et~al.}(2018)\citenamefont {Cavina},
  \citenamefont {Mancino}, \citenamefont {De~Pasquale}, \citenamefont
  {Gianani}, \citenamefont {Sbroscia}, \citenamefont {Booth}, \citenamefont
  {Roccia}, \citenamefont {Raimondi}, \citenamefont {Giovannetti},\ and\
  \citenamefont {Barbieri}}]{PhysRevA.98.050101}%
  \BibitemOpen
  \bibfield  {author} {\bibinfo {author} {\bibfnamefont {V.}~\bibnamefont
  {Cavina}}, \bibinfo {author} {\bibfnamefont {L.}~\bibnamefont {Mancino}},
  \bibinfo {author} {\bibfnamefont {A.}~\bibnamefont {De~Pasquale}}, \bibinfo
  {author} {\bibfnamefont {I.}~\bibnamefont {Gianani}}, \bibinfo {author}
  {\bibfnamefont {M.}~\bibnamefont {Sbroscia}}, \bibinfo {author}
  {\bibfnamefont {R.~I.}\ \bibnamefont {Booth}}, \bibinfo {author}
  {\bibfnamefont {E.}~\bibnamefont {Roccia}}, \bibinfo {author} {\bibfnamefont
  {R.}~\bibnamefont {Raimondi}}, \bibinfo {author} {\bibfnamefont
  {V.}~\bibnamefont {Giovannetti}},\ and\ \bibinfo {author} {\bibfnamefont
  {M.}~\bibnamefont {Barbieri}},\ }\bibfield  {title} {\bibinfo {title}
  {Bridging thermodynamics and metrology in nonequilibrium quantum
  thermometry},\ }\href {https://doi.org/10.1103/PhysRevA.98.050101} {\bibfield
   {journal} {\bibinfo  {journal} {Phys. Rev. A}\ }\textbf {\bibinfo {volume}
  {98}},\ \bibinfo {pages} {050101} (\bibinfo {year} {2018})}\BibitemShut
  {NoStop}%
\bibitem [{\citenamefont {Paris}(2016)}]{Paris_2015}%
  \BibitemOpen
  \bibfield  {author} {\bibinfo {author} {\bibfnamefont {M.~G.~A.}\
  \bibnamefont {Paris}},\ }\bibfield  {title} {\bibinfo {title} {Achieving the
  {L}andau bound to precision of quantum thermometry in systems with vanishing
  gap},\ }\href {https://doi.org/10.1088/1751-8113/49/3/03lt02} {\bibfield
  {journal} {\bibinfo  {journal} {J. Phys. A: Math. Theor.}\ }\textbf {\bibinfo
  {volume} {49}},\ \bibinfo {pages} {03LT02} (\bibinfo {year}
  {2016})}\BibitemShut {NoStop}%
\bibitem [{\citenamefont {Levy}\ and\ \citenamefont
  {Kosloff}(2014)}]{Levy_2014}%
  \BibitemOpen
  \bibfield  {author} {\bibinfo {author} {\bibfnamefont {A.}~\bibnamefont
  {Levy}}\ and\ \bibinfo {author} {\bibfnamefont {R.}~\bibnamefont {Kosloff}},\
  }\bibfield  {title} {\bibinfo {title} {The local approach to quantum
  transport may violate the second law of thermodynamics},\ }\href
  {https://doi.org/10.1209/0295-5075/107/20004} {\bibfield  {journal} {\bibinfo
   {journal} {Europhys. Lett.}\ }\textbf {\bibinfo {volume} {107}},\ \bibinfo
  {pages} {20004} (\bibinfo {year} {2014})}\BibitemShut {NoStop}%
\bibitem [{\citenamefont {Karg\ifmmode\imath\else\i\fi{}}\ \emph
  {et~al.}(2019)\citenamefont {Karg\ifmmode\imath\else\i\fi{}}, \citenamefont
  {Naseem}, \citenamefont {Opatrn\'y}, \citenamefont {M\"ustecapl\ifmmode
  \imath \else \i \fi{}o\ifmmode~\breve{g}\else \u{g}\fi{}lu},\ and\
  \citenamefont {Kurizki}}]{PhysRevE.99.042121}%
  \BibitemOpen
  \bibfield  {author} {\bibinfo {author} {\bibfnamefont {C.}~\bibnamefont
  {Karg\ifmmode\imath\else\i\fi{}}}, \bibinfo {author} {\bibfnamefont {M.~T.}\
  \bibnamefont {Naseem}}, \bibinfo {author} {\bibfnamefont {T.~c.}\
  \bibnamefont {Opatrn\'y}}, \bibinfo {author} {\bibfnamefont {O.~E.}\
  \bibnamefont {M\"ustecapl\ifmmode \imath \else \i
  \fi{}o\ifmmode~\breve{g}\else \u{g}\fi{}lu}},\ and\ \bibinfo {author}
  {\bibfnamefont {G.}~\bibnamefont {Kurizki}},\ }\bibfield  {title} {\bibinfo
  {title} {Quantum optical two-atom thermal diode},\ }\href
  {https://doi.org/10.1103/PhysRevE.99.042121} {\bibfield  {journal} {\bibinfo
  {journal} {Phys. Rev. E}\ }\textbf {\bibinfo {volume} {99}},\ \bibinfo
  {pages} {042121} (\bibinfo {year} {2019})}\BibitemShut {NoStop}%
\bibitem [{\citenamefont {Naseem}\ \emph
  {et~al.}(2020{\natexlab{a}})\citenamefont {Naseem}, \citenamefont {Misra},\
  and\ \citenamefont {M\"ustecaplio\ifmmode~\breve{g}\else
  \u{g}\fi{}lu}}]{Naseem_2020}%
  \BibitemOpen
  \bibfield  {author} {\bibinfo {author} {\bibfnamefont {M.~T.}\ \bibnamefont
  {Naseem}}, \bibinfo {author} {\bibfnamefont {A.}~\bibnamefont {Misra}},\ and\
  \bibinfo {author} {\bibfnamefont {O.~E.}\ \bibnamefont
  {M\"ustecaplio\ifmmode~\breve{g}\else \u{g}\fi{}lu}},\ }\bibfield  {title}
  {\bibinfo {title} {Two-body quantum absorption refrigerators with
  optomechanical-like interactions},\ }\href
  {https://doi.org/10.1088/2058-9565/ab8d89} {\bibfield  {journal} {\bibinfo
  {journal} {Quantum Sci. Technol.}\ }\textbf {\bibinfo {volume} {5}},\
  \bibinfo {pages} {035006} (\bibinfo {year} {2020}{\natexlab{a}})}\BibitemShut
  {NoStop}%
\bibitem [{\citenamefont {Naseem}\ \emph
  {et~al.}(2020{\natexlab{b}})\citenamefont {Naseem}, \citenamefont {Misra},
  \citenamefont {M\"ustecaplio\ifmmode~\breve{g}\else \u{g}\fi{}lu},\ and\
  \citenamefont {Kurizki}}]{PhysRevResearch.2.033285}%
  \BibitemOpen
  \bibfield  {author} {\bibinfo {author} {\bibfnamefont {M.~T.}\ \bibnamefont
  {Naseem}}, \bibinfo {author} {\bibfnamefont {A.}~\bibnamefont {Misra}},
  \bibinfo {author} {\bibfnamefont {O.~E.}\ \bibnamefont
  {M\"ustecaplio\ifmmode~\breve{g}\else \u{g}\fi{}lu}},\ and\ \bibinfo {author}
  {\bibfnamefont {G.}~\bibnamefont {Kurizki}},\ }\bibfield  {title} {\bibinfo
  {title} {Minimal quantum heat manager boosted by bath spectral filtering},\
  }\href {https://doi.org/10.1103/PhysRevResearch.2.033285} {\bibfield
  {journal} {\bibinfo  {journal} {Phys. Rev. Research}\ }\textbf {\bibinfo
  {volume} {2}},\ \bibinfo {pages} {033285} (\bibinfo {year}
  {2020}{\natexlab{b}})}\BibitemShut {NoStop}%
\bibitem [{\citenamefont {Kolář}\ \emph {et~al.}()\citenamefont {Kolář},
  \citenamefont {Guarnieri},\ and\ \citenamefont
  {Filip}}]{kolavr2022achieving}%
  \BibitemOpen
  \bibfield  {author} {\bibinfo {author} {\bibfnamefont {M.}~\bibnamefont
  {Kolář}}, \bibinfo {author} {\bibfnamefont {G.}~\bibnamefont {Guarnieri}},\
  and\ \bibinfo {author} {\bibfnamefont {R.}~\bibnamefont {Filip}},\
  }\href@noop {} {\bibinfo {title} {Achieving local coherence in thermal states
  by intra-system coupling}},\ \Eprint {https://arxiv.org/abs/2211.08851}
  {arXiv:2211.08851} \BibitemShut {NoStop}%
\bibitem [{\citenamefont {Breuer}\ and\ \citenamefont
  {Petruccione}(2007)}]{Breuer}%
  \BibitemOpen
  \bibfield  {author} {\bibinfo {author} {\bibfnamefont {H.-P.}\ \bibnamefont
  {Breuer}}\ and\ \bibinfo {author} {\bibfnamefont {F.}~\bibnamefont
  {Petruccione}},\ }\href
  {https://doi.org/10.1093/acprof:oso/9780199213900.001.0001} {\emph {\bibinfo
  {title} {{The Theory of Open Quantum Systems}}}}\ (\bibinfo  {publisher}
  {Oxford University Press, New York},\ \bibinfo {year} {2007})\BibitemShut
  {NoStop}%
\bibitem [{\citenamefont {Damanet}\ and\ \citenamefont
  {Martin}(2016)}]{Damanet_2016}%
  \BibitemOpen
  \bibfield  {author} {\bibinfo {author} {\bibfnamefont {F.}~\bibnamefont
  {Damanet}}\ and\ \bibinfo {author} {\bibfnamefont {J.}~\bibnamefont
  {Martin}},\ }\bibfield  {title} {\bibinfo {title} {Competition between
  finite-size effects and dipole{\textendash}dipole interactions in few-atom
  systems},\ }\href {https://doi.org/10.1088/0953-4075/49/22/225501} {\bibfield
   {journal} {\bibinfo  {journal} {J. Phys. B: At. Mol. Opt. Phys.}\ }\textbf
  {\bibinfo {volume} {49}},\ \bibinfo {pages} {225501} (\bibinfo {year}
  {2016})}\BibitemShut {NoStop}%
\bibitem [{\citenamefont {Latune}\ \emph {et~al.}(2019)\citenamefont {Latune},
  \citenamefont {Sinayskiy},\ and\ \citenamefont
  {Petruccione}}]{PhysRevA.99.052105}%
  \BibitemOpen
  \bibfield  {author} {\bibinfo {author} {\bibfnamefont {C.~L.}\ \bibnamefont
  {Latune}}, \bibinfo {author} {\bibfnamefont {I.}~\bibnamefont {Sinayskiy}},\
  and\ \bibinfo {author} {\bibfnamefont {F.}~\bibnamefont {Petruccione}},\
  }\bibfield  {title} {\bibinfo {title} {Energetic and entropic effects of
  bath-induced coherences},\ }\href
  {https://doi.org/10.1103/PhysRevA.99.052105} {\bibfield  {journal} {\bibinfo
  {journal} {Phys. Rev. A}\ }\textbf {\bibinfo {volume} {99}},\ \bibinfo
  {pages} {052105} (\bibinfo {year} {2019})}\BibitemShut {NoStop}%
\bibitem [{\citenamefont
  {\ifmmode\mbox{\c{C}}\else\c{C}\fi{}akmak}(2020)}]{PhysRevE.102.042111}%
  \BibitemOpen
  \bibfield  {author} {\bibinfo {author} {\bibfnamefont {B.}~\bibnamefont
  {\ifmmode\mbox{\c{C}}\else\c{C}\fi{}akmak}},\ }\bibfield  {title} {\bibinfo
  {title} {Ergotropy from coherences in an open quantum system},\ }\href
  {https://doi.org/10.1103/PhysRevE.102.042111} {\bibfield  {journal} {\bibinfo
   {journal} {Phys. Rev. E}\ }\textbf {\bibinfo {volume} {102}},\ \bibinfo
  {pages} {042111} (\bibinfo {year} {2020})}\BibitemShut {NoStop}%
\bibitem [{\citenamefont {Paris}(2009)}]{paris2009quantum}%
  \BibitemOpen
  \bibfield  {author} {\bibinfo {author} {\bibfnamefont {M.~G.~A.}\
  \bibnamefont {Paris}},\ }\bibfield  {title} {\bibinfo {title} {Quantum
  estimation for quantum technology},\ }\href
  {https://doi.org/10.1142/S0219749909004839} {\bibfield  {journal} {\bibinfo
  {journal} {Intl. J. Quant. Inf.}\ }\textbf {\bibinfo {volume} {07}},\
  \bibinfo {pages} {125} (\bibinfo {year} {2009})}\BibitemShut {NoStop}%
\bibitem [{\citenamefont {Cram{\'e}r}(1999)}]{cramer1999mathematical}%
  \BibitemOpen
  \bibfield  {author} {\bibinfo {author} {\bibfnamefont {H.}~\bibnamefont
  {Cram{\'e}r}},\ }\href {http://www.jstor.org/stable/j.ctt1bpm9r4} {\emph
  {\bibinfo {title} {Mathematical Methods of Statistics}}}\ (\bibinfo
  {publisher} {Princeton University Press},\ \bibinfo {year} {Princeton, NJ,
  1999})\BibitemShut {NoStop}%
\bibitem [{\citenamefont {Helstrom}(1969)}]{helstrom1969quantum}%
  \BibitemOpen
  \bibfield  {author} {\bibinfo {author} {\bibfnamefont {C.~W.}\ \bibnamefont
  {Helstrom}},\ }\bibfield  {title} {\bibinfo {title} {Quantum detection and
  estimation theory},\ }\href {https://doi.org/10.1007/BF01007479} {\bibfield
  {journal} {\bibinfo  {journal} {J. Stat. Phys.}\ }\textbf {\bibinfo {volume}
  {1}},\ \bibinfo {pages} {231} (\bibinfo {year} {1969})}\BibitemShut {NoStop}%
\bibitem [{\citenamefont {Braunstein}\ and\ \citenamefont
  {Caves}(1994)}]{PhysRevLett.72.3439}%
  \BibitemOpen
  \bibfield  {author} {\bibinfo {author} {\bibfnamefont {S.~L.}\ \bibnamefont
  {Braunstein}}\ and\ \bibinfo {author} {\bibfnamefont {C.~M.}\ \bibnamefont
  {Caves}},\ }\bibfield  {title} {\bibinfo {title} {Statistical distance and
  the geometry of quantum states},\ }\href
  {https://doi.org/10.1103/PhysRevLett.72.3439} {\bibfield  {journal} {\bibinfo
   {journal} {Phys. Rev. Lett.}\ }\textbf {\bibinfo {volume} {72}},\ \bibinfo
  {pages} {3439} (\bibinfo {year} {1994})}\BibitemShut {NoStop}%
\bibitem [{\citenamefont {Dittmann}(1999)}]{dittmann1999explicit}%
  \BibitemOpen
  \bibfield  {author} {\bibinfo {author} {\bibfnamefont {J.}~\bibnamefont
  {Dittmann}},\ }\bibfield  {title} {\bibinfo {title} {Explicit formulae for
  the {B}ures metric},\ }\href {https://doi.org/10.1088/0305-4470/32/14/007}
  {\bibfield  {journal} {\bibinfo  {journal} {J. Phys. A: Math. Gen.}\ }\textbf
  {\bibinfo {volume} {32}},\ \bibinfo {pages} {2663} (\bibinfo {year}
  {1999})}\BibitemShut {NoStop}%
\bibitem [{\citenamefont {Zhong}\ \emph {et~al.}(2013)\citenamefont {Zhong},
  \citenamefont {Sun}, \citenamefont {Ma}, \citenamefont {Wang},\ and\
  \citenamefont {Nori}}]{PhysRevA.87.022337}%
  \BibitemOpen
  \bibfield  {author} {\bibinfo {author} {\bibfnamefont {W.}~\bibnamefont
  {Zhong}}, \bibinfo {author} {\bibfnamefont {Z.}~\bibnamefont {Sun}}, \bibinfo
  {author} {\bibfnamefont {J.}~\bibnamefont {Ma}}, \bibinfo {author}
  {\bibfnamefont {X.}~\bibnamefont {Wang}},\ and\ \bibinfo {author}
  {\bibfnamefont {F.}~\bibnamefont {Nori}},\ }\bibfield  {title} {\bibinfo
  {title} {Fisher information under decoherence in {B}loch representation},\
  }\href {https://doi.org/10.1103/PhysRevA.87.022337} {\bibfield  {journal}
  {\bibinfo  {journal} {Phys. Rev. A}\ }\textbf {\bibinfo {volume} {87}},\
  \bibinfo {pages} {022337} (\bibinfo {year} {2013})}\BibitemShut {NoStop}%
\bibitem [{\citenamefont {Aspelmeyer}\ \emph {et~al.}(2014)\citenamefont
  {Aspelmeyer}, \citenamefont {Kippenberg},\ and\ \citenamefont
  {Marquardt}}]{RevModPhys.86.1391}%
  \BibitemOpen
  \bibfield  {author} {\bibinfo {author} {\bibfnamefont {M.}~\bibnamefont
  {Aspelmeyer}}, \bibinfo {author} {\bibfnamefont {T.~J.}\ \bibnamefont
  {Kippenberg}},\ and\ \bibinfo {author} {\bibfnamefont {F.}~\bibnamefont
  {Marquardt}},\ }\bibfield  {title} {\bibinfo {title} {Cavity optomechanics},\
  }\href {https://doi.org/10.1103/RevModPhys.86.1391} {\bibfield  {journal}
  {\bibinfo  {journal} {Rev. Mod. Phys.}\ }\textbf {\bibinfo {volume} {86}},\
  \bibinfo {pages} {1391} (\bibinfo {year} {2014})}\BibitemShut {NoStop}%
\bibitem [{\citenamefont {Massel}\ \emph {et~al.}(2012)\citenamefont {Massel},
  \citenamefont {Cho}, \citenamefont {Pirkkalainen}, \citenamefont {Hakonen},
  \citenamefont {Heikkil{\"a}},\ and\ \citenamefont
  {Sillanp{\"a}{\"a}}}]{Massel2012}%
  \BibitemOpen
  \bibfield  {author} {\bibinfo {author} {\bibfnamefont {F.}~\bibnamefont
  {Massel}}, \bibinfo {author} {\bibfnamefont {S.~U.}\ \bibnamefont {Cho}},
  \bibinfo {author} {\bibfnamefont {J.-M.}\ \bibnamefont {Pirkkalainen}},
  \bibinfo {author} {\bibfnamefont {P.~J.}\ \bibnamefont {Hakonen}}, \bibinfo
  {author} {\bibfnamefont {T.~T.}\ \bibnamefont {Heikkil{\"a}}},\ and\ \bibinfo
  {author} {\bibfnamefont {M.~A.}\ \bibnamefont {Sillanp{\"a}{\"a}}},\
  }\bibfield  {title} {\bibinfo {title} {Multimode circuit optomechanics near
  the quantum limit},\ }\href {https://doi.org/10.1038/ncomms1993} {\bibfield
  {journal} {\bibinfo  {journal} {Nat. Commun.}\ }\textbf {\bibinfo {volume}
  {3}},\ \bibinfo {pages} {987} (\bibinfo {year} {2012})}\BibitemShut {NoStop}%
\bibitem [{\citenamefont {Onofrio}(2016)}]{Onofrio_2016}%
  \BibitemOpen
  \bibfield  {author} {\bibinfo {author} {\bibfnamefont {R.}~\bibnamefont
  {Onofrio}},\ }\bibfield  {title} {\bibinfo {title} {Physics of our days:
  Cooling and thermometry of atomic {F}ermi gases},\ }\href
  {https://doi.org/10.3367/UFNe.2016.07.037873} {\bibfield  {journal} {\bibinfo
   {journal} {Physics-Uspekhi}\ }\textbf {\bibinfo {volume} {59}},\ \bibinfo
  {pages} {1129} (\bibinfo {year} {2016})}\BibitemShut {NoStop}%
\bibitem [{\citenamefont {Upadhyay}\ \emph {et~al.}(2021)\citenamefont
  {Upadhyay}, \citenamefont {Naseem}, \citenamefont {Marathe},\ and\
  \citenamefont {M\"ustecapl\ifmmode \imath \else \i
  \fi{}o\ifmmode~\breve{g}\else \u{g}\fi{}lu}}]{PhysRevE.104.054137}%
  \BibitemOpen
  \bibfield  {author} {\bibinfo {author} {\bibfnamefont {V.}~\bibnamefont
  {Upadhyay}}, \bibinfo {author} {\bibfnamefont {M.~T.}\ \bibnamefont
  {Naseem}}, \bibinfo {author} {\bibfnamefont {R.}~\bibnamefont {Marathe}},\
  and\ \bibinfo {author} {\bibfnamefont {O.~E.}\ \bibnamefont
  {M\"ustecapl\ifmmode \imath \else \i \fi{}o\ifmmode~\breve{g}\else
  \u{g}\fi{}lu}},\ }\bibfield  {title} {\bibinfo {title} {Heat rectification by
  two qubits coupled with {D}zyaloshinskii-{M}oriya interaction},\ }\href
  {https://doi.org/10.1103/PhysRevE.104.054137} {\bibfield  {journal} {\bibinfo
   {journal} {Phys. Rev. E}\ }\textbf {\bibinfo {volume} {104}},\ \bibinfo
  {pages} {054137} (\bibinfo {year} {2021})}\BibitemShut {NoStop}%
\bibitem [{\citenamefont {Law}(1994)}]{PhysRevA.49.433}%
  \BibitemOpen
  \bibfield  {author} {\bibinfo {author} {\bibfnamefont {C.~K.}\ \bibnamefont
  {Law}},\ }\bibfield  {title} {\bibinfo {title} {Effective {H}amiltonian for
  the radiation in a cavity with a moving mirror and a time-varying dielectric
  medium},\ }\href {https://doi.org/10.1103/PhysRevA.49.433} {\bibfield
  {journal} {\bibinfo  {journal} {Phys. Rev. A}\ }\textbf {\bibinfo {volume}
  {49}},\ \bibinfo {pages} {433} (\bibinfo {year} {1994})}\BibitemShut
  {NoStop}%
\bibitem [{\citenamefont {Law}(1995)}]{PhysRevA.51.2537}%
  \BibitemOpen
  \bibfield  {author} {\bibinfo {author} {\bibfnamefont {C.~K.}\ \bibnamefont
  {Law}},\ }\bibfield  {title} {\bibinfo {title} {Interaction between a moving
  mirror and radiation pressure: A {H}amiltonian formulation},\ }\href
  {https://doi.org/10.1103/PhysRevA.51.2537} {\bibfield  {journal} {\bibinfo
  {journal} {Phys. Rev. A}\ }\textbf {\bibinfo {volume} {51}},\ \bibinfo
  {pages} {2537} (\bibinfo {year} {1995})}\BibitemShut {NoStop}%
\bibitem [{\citenamefont {Moqadam}\ \emph {et~al.}(2015)\citenamefont
  {Moqadam}, \citenamefont {Portugal},\ and\ \citenamefont
  {de~Oliveira}}]{Moqadam2015}%
  \BibitemOpen
  \bibfield  {author} {\bibinfo {author} {\bibfnamefont {J.~K.}\ \bibnamefont
  {Moqadam}}, \bibinfo {author} {\bibfnamefont {R.}~\bibnamefont {Portugal}},\
  and\ \bibinfo {author} {\bibfnamefont {M.~C.}\ \bibnamefont {de~Oliveira}},\
  }\bibfield  {title} {\bibinfo {title} {Quantum walks on a circle with
  optomechanical systems},\ }\href {https://doi.org/10.1007/s11128-015-1079-9}
  {\bibfield  {journal} {\bibinfo  {journal} {Quantum Inf Process}\ }\textbf
  {\bibinfo {volume} {14}},\ \bibinfo {pages} {3595} (\bibinfo {year}
  {2015})}\BibitemShut {NoStop}%
\bibitem [{\citenamefont {Phoenix}\ and\ \citenamefont
  {Knight}(1990)}]{Phoenix1990}%
  \BibitemOpen
  \bibfield  {author} {\bibinfo {author} {\bibfnamefont {S.~J.~D.}\
  \bibnamefont {Phoenix}}\ and\ \bibinfo {author} {\bibfnamefont {P.~L.}\
  \bibnamefont {Knight}},\ }\bibfield  {title} {\bibinfo {title} {Periodicity,
  phase, and entropy in models of two-photonresonance},\ }\href
  {https://doi.org/10.1364/JOSAB.7.000116} {\bibfield  {journal} {\bibinfo
  {journal} {J. Opt. Soc. Am. B}\ }\textbf {\bibinfo {volume} {7}},\ \bibinfo
  {pages} {116} (\bibinfo {year} {1990})}\BibitemShut {NoStop}%
\bibitem [{\citenamefont {Blais}\ \emph {et~al.}(2004)\citenamefont {Blais},
  \citenamefont {Huang}, \citenamefont {Wallraff}, \citenamefont {Girvin},\
  and\ \citenamefont {Schoelkopf}}]{PhysRevA.69.062320}%
  \BibitemOpen
  \bibfield  {author} {\bibinfo {author} {\bibfnamefont {A.}~\bibnamefont
  {Blais}}, \bibinfo {author} {\bibfnamefont {R.-S.}\ \bibnamefont {Huang}},
  \bibinfo {author} {\bibfnamefont {A.}~\bibnamefont {Wallraff}}, \bibinfo
  {author} {\bibfnamefont {S.~M.}\ \bibnamefont {Girvin}},\ and\ \bibinfo
  {author} {\bibfnamefont {R.~J.}\ \bibnamefont {Schoelkopf}},\ }\bibfield
  {title} {\bibinfo {title} {Cavity quantum electrodynamics for superconducting
  electrical circuits: An architecture for quantum computation},\ }\href
  {https://doi.org/10.1103/PhysRevA.69.062320} {\bibfield  {journal} {\bibinfo
  {journal} {Phys. Rev. A}\ }\textbf {\bibinfo {volume} {69}},\ \bibinfo
  {pages} {062320} (\bibinfo {year} {2004})}\BibitemShut {NoStop}%
\bibitem [{\citenamefont {Travaglione}\ and\ \citenamefont
  {Milburn}(2002)}]{PhysRevA.65.032310}%
  \BibitemOpen
  \bibfield  {author} {\bibinfo {author} {\bibfnamefont {B.~C.}\ \bibnamefont
  {Travaglione}}\ and\ \bibinfo {author} {\bibfnamefont {G.~J.}\ \bibnamefont
  {Milburn}},\ }\bibfield  {title} {\bibinfo {title} {Implementing the quantum
  random walk},\ }\href {https://doi.org/10.1103/PhysRevA.65.032310} {\bibfield
   {journal} {\bibinfo  {journal} {Phys. Rev. A}\ }\textbf {\bibinfo {volume}
  {65}},\ \bibinfo {pages} {032310} (\bibinfo {year} {2002})}\BibitemShut
  {NoStop}%
\end{thebibliography}%
%--------------------------------------------------------------------
\end{document}